\documentclass[11pt,a4paper]{article}
\pdfoutput=1

\usepackage{jheppub}
\usepackage{latexsym}
\usepackage{multirow}
\usepackage{color}
\usepackage[usenames,dvipsnames,table]{xcolor}
\usepackage{graphicx}
\usepackage{epsfig} 
\usepackage{epsf}   
\usepackage{dcolumn}
\usepackage{bm}
\usepackage{dcolumn}
\usepackage{textcomp}
\usepackage{float}
\usepackage{subfig}
\usepackage{hypcap}
\usepackage[]{hyperref}
\usepackage{makecell}
\usepackage{color}
\usepackage{pifont}
\usepackage{appendix}
\usepackage{amsmath}
\usepackage{multirow,bigdelim}  
\usepackage{lineno}
\usepackage[utf8x]{inputenc}
\usepackage[normalem]{ulem}
\graphicspath{{fig/}}

\newcommand{\comment}[1]{}
\usepackage{amssymb}
\hypersetup{
	bookmarks=true,         
	unicode=false,          
	pdftoolbar=true,        
	pdfmenubar=true,        
	pdffitwindow=true,   
	pdfstartview={FitH},    
	pdfsubject={scalar NSI},   
	pdfnewwindow=true,      
	pdfcreator={RevTeX},
	colorlinks=true,     
	linkcolor=red,         
	citecolor=blue,        
	filecolor=black,     
	urlcolor=blue,           
}

\preprint{}

\title{The Sensitivity of DUNE in Presence of Off-Diagonal Scalar NSI Parameters}

\author[a,1]{Arnab Sarker,} 
\author[a,2]{Dharitree Bezboruah,}
\author[b,3]{Abinash Medhi,}
\author[a,4]{Moon Moon Devi,}

\affiliation[a]{Department of Physics, Tezpur University, Napaam, Sonitpur, Assam-784028, India}
\affiliation[b]{Department of Physics, Indian Institute of Technology, Guwahati, Assam-781039, India}

\emailAdd{$^{1}$arnabs@tezu.ernet.in}
\emailAdd{$^{2}$dbbphy1@tezu.ernet.in}
\emailAdd{$^{3}$abinashmedhi0@gmail.com}
\emailAdd{$^{4}$devimm@tezu.ernet.in}

\date{\today}

\abstract{Scalar non-standard interactions (NSI) presents an exciting pathway for probing potential new physics that extends beyond the Standard Model (BSM). The scalar coupling of neutrinos with matter can appear as a sub-dominant effect that can impact the neutrino oscillation probabilities. The uniqueness of these interactions is that it can directly affect the neutrino mass matrix. This makes oscillations sensitive to the absolute neutrino mass. The effects of scalar NSI scales linearly with matter density which motivates its exploration in long-baseline sector. The presence of scalar NSI can influence the key measurements in the field of neutrino physics, including the precise determination of the leptonic CP phase ($\delta_{CP}$), neutrino mass ordering and the octant of $\theta_{23}$. The precise determination of $\delta_{CP}$ is one of the major goals of DUNE, which is an upcoming long-baseline experiment. A better understanding of the impact of scalar NSI on CP measurement sensitivities is crucial for accurate interpretation of $\delta_{CP}$ phase. In this work, we have explored the impact of the complex off-diagonal scalar NSI elements $\eta_{\alpha\beta}$ and their associated phases $\phi_{\alpha\beta}$ on the CP-measurement sensitivities at DUNE. We have explored the impact of the neutrino mass scale on these sensitivities. We look for constraining these off-diagonal elements for different neutrino mass scales. We also explore their correlation with $\delta_{CP}$, investigating potential degeneracies that can arise due to additional phases. We also perform a correlation study among different scalar NSI elements. We show that the inclusion of the complex scalar NSI elements can significantly modify the CP phase measurements.}


\begin{document}
\maketitle

\section{Introduction} \label{sec:introduction}

Neutrino physics has undergone significant advancements over the past few decades, with substantial progress made in understanding the properties of neutrinos. The excellent ability of the Standard Model (SM) to explain the phenomena around us makes it evident that SM is an effective theory and that any new physics beyond SM (BSM) is suppressed at low energies. The experimental confirmation of neutrino oscillations implies that neutrinos ($\nu's$) have mass and that neutrino flavors can change in the course of their propagation. This was a milestone that reshaped our understanding and provided evidence for physics beyond the Standard Model. The neutrinos are described in the standard 3$\nu$ paradigm, with three active neutrino flavors ($\nu_e, \nu_\mu, \nu_\tau$) defined as the superposition of three definite mass eigenstates ($\nu_1$, $\nu_2$, $\nu_3$). The production, propagation and detection of these neutrinos are solely mediated by weak interactions.

With high-precision accelerator experiments like T2K \cite{T2K:2014xyt} and NO$\nu$A \cite{NOvA:2004blv, NOvA:2016kwd}, neutrino oscillation physics is at the dawn of precision measurement era, allowing a robust test of the 3$\nu$ paradigm. Most of the neutrino oscillation parameters are very precisely measured, in particular, $|\Delta m_{31}^{2}|$, $\Delta m_{21}^{2}$, $\theta_{12}$ and $\theta_{13}$. However, there exist three unknowns i.e. the leptonic CP-violating phase ($\delta_{CP}$), the sign of $\Delta m_{31}^{2}$ and the octant of $\theta_{23}$. Driven by significant advancements in beam technology, detector capabilities and analysis methods, the upcoming long-baseline (LBL) experiments like DUNE \cite{DUNE:2015lol}, HK \cite{ Hyper-KamiokandeProto-:2015xww, Hyper-Kamiokande:2016srs} and P2SO \cite{Akindinov:2019flp} will have the capability to solve these three key unknowns remaining in the field of neutrino physics. Currently, T2K data has shown a preference for maximal CP violation with $\delta_{CP} \sim 3\pi /2$ \cite{T2K:2023smv}, whereas NO$\nu$A data has no preference for either CP violation or conservation \cite{NOvA:2021nfi}. The tension in T2K and NO$\nu$A data can be the indication of new physics beyond non-zero neutrino mass. In order to quantify leptonic CP-violation (CPV), a precise measurement of all the other oscillation parameters is necessary. Many studies on the impact of precision measurements and their interplay have been done as shown in references \cite{Coloma:2012wq,Minakata:2013eoa,Coloma:2014kca,Agarwalla:2022xdo,Denton:2023zwa}.

While the three-flavor oscillation framework has been remarkably successful in explaining most of neutrino data, there is growing interest in exploring potential deviations from this paradigm that could arise from new physics scenarios. With the increasing sensitivity of the detectors with state-of-the-art technologies, it has now become crucial to probe and understand such subdominant new physics scenarios. In the neutrino sector, the new physics effects include the possibility of more than three neutrino states, neutrino decay \cite{Berryman:2014qha,PICORETI201670,SNO:2018pvg,GOMES2015345,Coloma:2017zpg,Abrahao:2015rba}, non-standard interactions (NSIs) \cite{Liao:2016orc,Friedland:2012tq,Coelho:2012bp,Rahman:2015vqa,Coloma:2015kiu,deGouvea:2015ndi,Liao:2016hsa,Forero:2016cmb,Huitu:2016bmb,Bakhti:2016prn,Kumar:2021lrn,Agarwalla:2015cta,Agarwalla:2014bsa,Agarwalla:2012wf,Blennow:2016etl,Blennow:2015nxa,Deepthi:2016erc,Masud:2021ves,Soumya:2019kto,Masud:2018pig,Masud:2017kdi,Masud:2015xva,Ge:2016dlx,Fukasawa:2016lew,Chatterjee:2021wac,Medhi:2023ebi,Chaves:2021kxe,Brahma:2023wlf,Davoudiasl:2023uiq,Chatterjee:2020kkm,Choubey:2014iia,Singha:2021jkn,Denton:2018xmq,Denton:2020uda,Farzan:2015hkd,Ge:2018uhz, Medhi:2021wxj}, Lorentz Invariance Violation (LIV) \cite{Kostelecky:2003cr, SNO:2018mge,Mewes:2019dhj, Huang:2019etr, ARIAS2007401,LSND:2005oop,MINOS:2008fnv,MINOS:2010kat,IceCube:2010fyu,MiniBooNE:2011pix,DoubleChooz:2012eiq, Sarker:2023mlz,Sarkar:2022ujy,Majhi:2022fed}, quantum decoherence \cite{PhysRevD.56.6648,Benatti:2000ph,PhysRevD.100.055023,PhysRevD.95.113005,Lisi:2000zt} etc. 
Wolfenstein first proposed the concept of NSI with generally parameterized vector and axial vector currents in his seminal paper \cite{Wolfenstein:1977ue} on matter effect in neutrinos. The vector-mediated NSI can generically be of CC type which affects the production, propagation and detection of neutrinos or NC type which affects only production and detection. In last two decades, both theoretical and phenomenological exploration of vector NSI have been done extensively in the literature. The presence of NSI introduces various degeneracies in oscillation and scattering experiments, which may highly affect the precision measurement of neutrino oscillation parameters \cite{Miranda:2004nb,Gonzalez-Garcia:2001snt,Kikuchi:2008vq,Friedland:2012tq,Rahman:2015vqa,Masud:2015xva,Coloma:2015kiu,Palazzo:2015gja,deGouvea:2015ndi,Masud:2016bvp,Liao:2016hsa,deGouvea:2016pom,Liao:2016orc,Ge:2016dlx,Agarwalla:2016fkh,Blennow:2016etl,Fukasawa:2016lew,Deepthi:2016erc,Forero:2016cmb,Deepthi:2017gxg,Coloma:2017egw,Coloma:2017ncl,Hyde:2018tqt,Esteban:2019lfo,Kopp:2010qt}. This makes the complementary measurement of NSI parameters crucial for the robust interpretation of data from future sophisticated experiments.

One of the primary goals of current and future neutrino experiments is the precise measurement of the $\delta_{CP}$ phase. This measurement is crucial for understanding the matter-antimatter asymmetry in the universe, as CP-violation in the lepton sector could potentially contribute to the observed dominance of matter over antimatter. A non-zero value of the CP phase in the Pontecorvo-Maki-Nakagawa-Sakata (PNMS) \cite{Pontecorvo:1957cp,Pontecorvo:1957qd,Pontecorvo:1967fh,ParticleDataGroup:2020ssz} matrix is indispensable for a viable explanation of the asymmetry observed in the universe based on leptogenesis \cite{Fukugita:1986hr, Davidson:2008bu}. The appearance probability channel in accelerator-based experiments is the most straightforward way to investigate leptonic CPV \cite{T2K:2001wmr,Marciano:2001tz,Dick:1999ed,Cervera:2000kp,Nunokawa:2007qh}.
In measuring the leptonic CP phase, the presence of NSI can have a huge impact as they may introduce new sources of CP-violation apart from the Dirac CP phase. The presence of new CP-violating phases may alter the relationship between the measured oscillation parameters and the Dirac CP phase \cite{Gonzalez-Garcia:2001snt}. In the standard scenario, DUNE has the potential to observe CP-violation at 5$\sigma$ confidence level (CL) after 10 years of data taking for more than 50$\%$ of the $\delta_{CP}$ values \cite{DUNE:2020jqi}. The presence of vector NSI can spoil the CPV sensitivities at DUNE as shown in \cite{Masud:2016bvp}. The NSI-SI degeneracies and the interplay of their moduli and phases have been explored in \cite{Masud:2015xva} for precise CP phase measurement. The reference \cite{Rout:2017udo} explores the impact of new physics scenarios on distinguishing between intrinsic and extrinsic CP-violation at LBL experiments. However, by exploiting the flexibility of the tunable beam line in DUNE, better discrimination of NSI from SI can be achieved, as presented in \cite{Masud:2017bcf}. In reference \cite{Ghosh:2013pfa}, it is shown that synergy among different experiments can lift the $\delta_{CP}$-hierarchy degeneracy and hence provide better sensitivity towards CP-violation. The authors in reference \cite{Choubey:2017cba} have studied the sensitivity of DUNE+T2HKK towards $\delta_{CP}$ measurement in the presence of one sterile neutrino. In reference \cite{Prakash:2012az}, it is shown that combining data from the T2K and NO$\nu$A could break the degeneracy between neutrino mass hierarchies and $\delta_{CP}$. In a recent study \cite{Agarwalla:2022xdo}, authors have explored the possibility of enhanced CP-violation sensitivity by combining data from DUNE, T2HK and T2HKK. In reference \cite{Denton:2023qmd}, the authors have suggested the exploration of the disappearance channel for determining CP-violation which will help in the robustness of $\delta_{CP}$ measurement.

Beyond the well-established vector NSI, scalar field-mediated NSI \cite{Ge:2018uhz}, presents a fascinating avenue for exploring new physics beyond the Standard Model. Unlike vector NSI, scalar NSI directly modifies the neutrino masses through their Yukawa-like terms in the Lagrangian. This results in a distinct phenomenological signature in neutrino experiments, as shown by various recent studies. In references \cite{Medhi:2021wxj,Medhi:2022qmu}, the authors have explored the CPV sensitivities at DUNE considering only the diagonal scalar NSI elements and performed a synergy study of DUNE, T2HK and T2HKK. A similar study is also performed for DUNE and P2SO with diagonal scalar NSI elements for CPV and octant sensitivities in \cite{Singha:2023set}. They have also explored the constraining capabilities of DUNE and P2SO towards these diagonal scalar NSI parameters. The authors in reference \cite{Medhi:2023ebi} have explored the possibility of constraining the absolute neutrino mass in the presence of scalar NSI at DUNE. The impact on the neutrino mass ordering is studied in reference \cite{Sarker:2023qzp} in the context of LBL experiments i.e. DUNE, HK and HK+KNO. The authors in reference \cite{Denton:2022pxt} have explored the possibility of distinguishing between different new physics scenarios, such as the presence of sterile neutrinos, vector and scalar NSI at NO$\nu$A, T2K and DUNE. In reference \cite{ESSnuSB:2023lbg}, the effect of scalar NSI on $\delta_{CP}$ measurement is studied in the context of the ESSnuSB experiment. The reference \cite{Chaves:2021kxe} has explored scalar NSI using $\nu$($\bar{\nu}$)-disappearance in oscillation data at different experiments. The studies like \cite{Babu:2019iml,Venzor:2020ova} have set bounds on scalar NSI within cosmological and astrophysical limits, though these constraints are relatively loose. The authors of \cite{Gupta:2023wct} have constrained scalar NSI parameters considering the JUNO experiment. In reference \cite{Dutta:2024hqq}, the authors have explored the sensitivities at DUNE and compared DUNE's constraint with other non-oscillatory probes.

In this work, we build upon these efforts by conducting a comprehensive study for the first time on the effects of off-diagonal scalar NSI parameters  $\eta_{\alpha\beta}$ on the CP-measurement sensitivities at DUNE. We perform a model-independent analysis, exploring the impact of the off-diagonal elements $\eta_{\alpha\beta}$ and their associated phases $\phi_{\alpha\beta}$ on the $\nu$-oscillation probabilities. We will particularly explore the impact of new additional CP phases in the presence of scalar NSI on the CP-measurement sensitivities at DUNE. The presence of additional phases can lead to degeneracy in the $\delta_{CP}$ measurement. Hence, it is necessary to study the impact of these phases on the CP-measurement. This is particularly important as it can introduce more complexities and degeneracies in the CP measurements. A better understanding of the impact is essential for accurately isolating the intrinsic CP-violating phase. This study will shed light on the impact of additional phases due to scalar NSI and the reliability of CP phase measurements at upcoming long-baseline neutrino experiments i.e. DUNE. We will also set limits on the off-diagonal elements at a given statistical significance. We will explore the correlation between different scalar NSI elements that may lead to possible degeneracies.

This article is organised as follows. In section \ref{sec:scalar_NSI}, we cover the theoretical background of scalar non-standard interactions. We then provide details of the technical aspects of DUNE and the simulation framework in section \ref{sec:methdology}. The impact of off-diagonal scalar NSI elements $\eta_{\alpha\beta}$ and $\phi_{\alpha\beta}$ at the probability level is explored in section \ref{sec:Pmue}. In section \ref{sec:CPmeasurement}, we examine how scalar NSI affects the CP-measurement sensitivities at DUNE. We particularly explore the impact of additional scalar NSI phases on the sensitivities. The bounds on the off-diagonal elements are presented in section \ref{sec:eta_bound}, and the correlation between different scalar NSI elements is discussed in section \ref{sec:corr}. We then summarize our findings in section \ref{sec:summary}.

\section{Theoretical background} \label{sec:scalar_NSI}
The interaction of neutrinos with fermions via a scalar-mediated particle can have interesting phenomenological consequences. In the presence of scalar non-standard interactions, the effective Lagrangian can be written as \cite{Wolfenstein:1977ue, Nieves:2003in, Nishi:2004st, Maki:1962mu, Bilenky:1987ty}

\begin{equation} \label{eq:scalar_lag}
  \mathcal{L}_{eff}^s = \frac{y_f y_{\alpha \beta}}{m_\phi^2} [\bar \nu_{\alpha} (p_3) \nu_\beta (p_2)] [\bar f(p_1) f(p_4)],
\end{equation}
where $y_{\alpha \beta}$ and $y_f$ represents the coupling of scalar mediator $\phi$ with neutrinos ($\nu_{\alpha,\beta}$ with $\alpha,\beta$ = $e,\mu,\tau$) and fermions/anti-fermions ($f$/$\bar f$). The presence of scalar NSI leads to a modification in the Dirac equation as shown in equation \ref{eq:Dirac_snsi}. The quantity $n_{f}$ represents the number density of matter fermions.

\begin{equation} \label{eq:Dirac_snsi}
 \bar \nu_\beta \left[ i \partial_\mu \gamma^\mu + \left( M_{\beta \alpha}   + \frac{ \sum_f n_f y_f y_{\alpha \beta}}{m_\phi^2} \right) \right] \nu_\alpha =0
\end{equation}

\noindent The Hamiltonian for standard neutrino-matter interactions in the flavor basis can be written as,
\begin{equation} \label{eq:Hamiltonian_SI}
\mathcal{H}_{matter}= E_\nu + \frac{M M^\dagger}{2 E_\nu} \pm V_{SI}
\end{equation}
Here, M represents the neutrino mass matrix in the flavor basis and $\mathcal{U}$ represents the PMNS neutrino mixing matrix. The equation \ref{eq:Hamiltonian_SI} can also be expressed as,
\begin{equation} \label{eq:Hamiltonian_SI_eff}
\mathcal{H}_{eff}= E_\nu + \frac{1}{2 E_\nu} \mathcal{U}.diag( 0, \Delta m_{21}^2, \Delta m_{31}^2).\mathcal{U}^\dagger + diag(V_{CC} ,0,0).
\end{equation}
\noindent As seen in equation \ref{eq:Dirac_snsi}, the contribution of scalar NSI arises as an addition to the neutrino mass matrix. The uniqueness of scalar NSI is that it can mimic the dependence on the true neutrino mass matrix. In the presence of scalar NSI, the Hamiltonian governing the $\nu$-oscillations can be rewritten in the form given below as, 

\begin{equation}\label{eq:Heff_SNSI}
   \mathcal{H}_{SNSI} \equiv E_\nu + \frac{M_{eff} M_{eff}^\dagger}{2 E_\nu} \pm V_{SI}.
\end{equation}

\noindent Here, $M_{eff} = M + \delta M$, with $\delta M \equiv \frac{\sum_f n_f y_f y_{\alpha \beta}}{m_{\phi}^2}$. We have followed the parameterization of $\delta M$ as shown in the references \cite{Ge:2018uhz,Medhi:2021wxj,Medhi:2022qmu} and the form of the matrix is shown in equation \ref{eq:deltaM_para}, where $\eta_{\alpha \beta}$ are dimensionless quantities. 

\begin{equation} \label{eq:deltaM_para}
\delta M\equiv S_{m}\left(\begin{array}{ccc}
\eta_{ee} & \eta_{e\mu} & \eta_{e\tau}\\
\eta_{e \mu}^* & \eta_{\mu\mu} & \eta_{\mu\tau}\\
\eta_{e \tau}^* & \eta_{\mu\tau}^* & \eta_{\tau\tau}
\end{array}\right)
\end{equation}

\noindent The parameters $\eta_{\alpha\beta}$ quantify the strength of scalar NSI and the scaling factor $S_{m}$ is merely introduced to re-express $\delta M$ in mass dimensions. In our work, the scaling factor is fixed at $S_{m} = \sqrt{|\Delta m_{31}^{2}|}$ which corresponds to the magnitude of a typical atmospheric mass-squared splitting. The Hamiltonian is hermitian in nature, which allows the diagonal parameters to be real and the off-diagonal parameters to be complex, as $\eta_{\alpha\beta}=|\eta_{\alpha\beta}|e^{-i\phi_{\alpha\beta}}$. Here, $|\eta_{\alpha\beta}|$ are the magnitudes of the off-diagonal scalar NSI elements, and $\phi_{\alpha\beta}$ are the corresponding phases. 

In the presence of scalar NSI, the neutrino oscillation probabilities will not only depend on the mass-squared differences but also on the absolute neutrino masses. As scalar NSI scales linearly with matter density, we have defined the values of scalar NSI parameters at matter density $\rho=2.9g/cm^{3}$ in this work. Considering the long-baseline sector, the impact of rescaling $\eta_{\alpha\beta}$ for varying density is nominal. The presence of off-diagonal phases $\phi_{\alpha\beta}$ can play an important role in the measurement of the Dirac CP phase $\delta_{CP}$. The phases $\phi_{\alpha\beta}$ may significantly modify the $\nu$-oscillation probabilities. It may also affect DUNE's sensitivity towards CP measurement. This is particularly significant as it can introduce additional complexities and degeneracies in CP-measurements. A deeper understanding of these impacts is crucial for accurately isolating the intrinsic CP-violating phase. This study will illuminate the effects of additional phases due to scalar NSI, enhancing the reliability of CP phase measurements at upcoming long-baseline neutrino experiments such as DUNE. In this study, we have investigated the effects of off-diagonal elements and their associated phases on DUNE's sensitivity to CP-violation.

\section{Experimental details and Numerical Analysis}\label{sec:methdology}

We first describe the specifics of the experimental configurations of DUNE and then elaborate on the details of the simulation procedure. Our study has specifically addressed the Deep Underground Neutrino Experiment (DUNE) \cite{DUNE:2016hlj, DUNE:2015lol, DUNE:2016rla, DUNE:2020ypp, DUNE:2021tad}, which is a leading accelerator-based $\nu$-experiment set to be hosted at Fermilab. Long-Baseline Neutrino Facility (LBNF) at Fermilab is capable of providing $1.1 \times 10^{21}$ protons on target (POT) annually through its 1.2 MW proton beam, which will generate the neutrinos for the experiment. The project encompasses two detectors, with the near detector located 60 meters underground at Fermilab, with a baseline of 574 meters. Meanwhile, the far detector, with a baseline of 1300 km, will be positioned 1.5 km beneath the Earth's surface at the Sanford Underground Research Facility in South Dakota. The far detector module will have four Liquid Argon Time Projection Chambers (LArTPC), each with a fiducial mass of 10 kt. The cutting-edge technologies will deliver outstanding capabilities for energy reconstruction, particle identification and tracking, enabling unparalleled precision in the study of neutrino properties. The details of detection channels and systematic uncertainties for DUNE are described in table \ref{tab:experimental_setup}.

\begin{table}[!t]
    \label{tab:detector_details}
    \centering
    \begin{tabular}{|c|c|c|c|}
    \hline
       \multirow{2}{*}{Experimental Detail } &\multirow{2}{*}{Detection Channels} & \multicolumn{2}{|c|}{Normalization error} \\\cline{3-4}
        & & Signal & Background \\
        \hline
         DUNE, Baseline = 1300 km  &  & & \\
         L/E = 1543 km/GeV &  $\nu_e (\bar \nu_e)$ appearance & 2 \% (2\%) & 5 \% (5 \%) \\
         Fiducial mass = 40 kt (LArTPC)& $\nu_\mu (\bar \nu_\mu)$ disappearance & 5 \% (5 \%) & 5 \% (5 \%) \\ 
         Runtime = 3.5 yr $\nu + 3.5$ yr $\bar \nu$  &  & & \\
    \hline
    \end{tabular}
    \caption{Detection channels and systematic uncertainties for DUNE.}
    \label{tab:experimental_setup}
\end{table}

We simulate the DUNE experiment using the General Long Baseline Experiment Simulator (GLoBES) \cite{Huber:2004ka, Kopp:2006wp, Huber:2007ji}; a sophisticated C-based framework used for simulating long-baseline neutrino oscillation experiments. The numerical calculation of low-level information, i.e., event rates and oscillation probabilities is possible with GLoBES. In addition, the framework also allows the extraction of statistical $\Delta \chi^{2}$ for any simulated long-baseline experiment. The computation of statistical $\Delta \chi^{2}$ helps in the sensitivity studies of long-baseline experiments. In our work, the statistical significance for the sensitivity studies considering the DUNE experiment is extracted from the GLoBES framework. The statistical $\chi^2$ for the distinction between the standard interaction and the effects due to scalar non-standard interactions can be defined as,
\begin{equation}
\label{eq:chisq}
\chi^2 \equiv  \min_{\eta}  \sum_{i} \sum_{j}
\frac{\left[N_{true}^{i,j} - N_{test}^{i,j} \right]^2 }{N_{true}^{i,j}},
\end{equation}

\begin{equation} \label{eq:chi2}
    \chi_{pull}^{2}=\underset{\zeta_{j}}{min}\left(\chi^{2}+\sum_{i=1}^{k}\frac{\zeta_{i}^{2}}{\sigma_{\zeta_{i}}^{2}}\right),
\end{equation}
The variables $N_{true}^{i,j}$ and $N_{test}^{i,j}$ indicate the count of true events and test events, respectively. We have included the systematic errors using the pull method described in \cite{Huber:2004ka,Fogli:2002pt}. To account for these errors, we introduced nuisance parameters represented by $\zeta_{k}$, where the systematic uncertainties associated with $\zeta_{k}$ are denoted by $\sigma_{\zeta_{k}}^{2}$. The overall statistical significance involves minimization over all the considered systematic errors, as shown in equation \ref{eq:chi2}.

We have modified the GLoBES simulation framework accordingly to incorporate the effects of scalar NSI. We have used the effective neutrino Hamiltonian in the presence of scalar NSI as shown in equation \ref{eq:Heff_SNSI}. The true values of the standard neutrino oscillation parameters are fixed at values as summarized in table \ref{tab:param}. We have also marginalized over the oscillation parameters $\theta_{23}$, $\Delta m_{31}^{2}$ and $\delta_{CP}$ in the range shown in table \ref{tab:param}. The neutrino mass ordering is considered to be known and fixed to true normal mass ordering (NO). The marginalization of the scalar NSI parameters $|\eta_{\alpha\beta}|$ and $\phi_{\alpha\beta}$ has also been performed wherever necessary. In the complete analysis, we have defined the values of scalar NSI $\eta_{\alpha\beta}$ at matter density $\rho=2.9g/cm^{3}$. For $\delta_{CP}$-measurement studies, we consider only one non-zero off-diagonal scalar NSI parameter at a time for the statistical analysis. However, in the correlation studies, we have considered two non-zero scalar NSI parameters at a time. The scalar NSI introduces a dependence of neutrino oscillations on the absolute neutrino masses. Therefore, in this study, we set the lightest neutrino mass at $10^{-5}eV$.

We first study the impact of off-diagonal scalar NSI elements ($|\eta_{e\mu}|$, $|\eta_{e\tau}|$, $|\eta_{\mu\tau}|$) and their corresponding phases ($\phi_{e\mu}$, $\phi_{e\tau}$, $\phi_{\mu\tau}$) on the neutrino oscillation probabilities in section \ref{sec:Pmue}. We particularly focus on the dominant oscillation channel, i.e., $P_{\mu e}$ in order to explore the effects at the probability level. We consider only one non-zero $\eta_{\alpha\beta}$ element at a time to understand the role of each individual off-diagonal parameter unless otherwise mentioned. We also explore DUNE's sensitivity towards simultaneous measurement of $\delta_{CP}$ and $\eta_{\alpha\beta}$ in section \ref{sec:dcp_precision}. We study the impact of $\eta_{\alpha\beta}$ and $\phi_{\alpha\beta}$ on the CP-violation sensitivities at DUNE in section \ref{sec:CPVsens}. The DUNE's capability to constrain the off-diagonal scalar NSI parameters is explored in section \ref{sec:eta_bound}. We finally explore the correlation among different scalar NSI elements considering the DUNE experiment in section \ref{sec:corr}.

\begin{table}[!t]
    \centering
    \begin{tabular}{|c|c|c|c|c|c|c|}
    \hline 
    Parameters & $\theta_{12}$ $[^{\circ}]$ & $\theta_{13}$ $[^{\circ}]$ & $\theta_{23}$$[^{\circ}]$ & $\Delta m_{21}^{2}$$[10^{-3}eV^{2}]$ & $\Delta m_{31}^{2}$$[10^{-5}eV^{2}]$ & $\delta_{CP} $\tabularnewline 
    \hline 
    Values & 33.45 & 8.62 & 47 & 7.42  & 2.55 & $-\pi/2$\tabularnewline
    Marginalization & Fixed & Fixed & 40-50 & Fixed & 2.25-2.65 & [$-\pi$, $\pi$]\tabularnewline 
    \hline 
    \end{tabular}
    \caption{Neutrino oscillation parameters are presented with their corresponding marginalization range \cite{Esteban:2020cvm}.}
    \label{tab:param}
\end{table}


\section{Exploration at probability level with $P_{\mu e}$}\label{sec:Pmue}
\begin{figure}[!h]
	\centering
    \includegraphics[width=0.328\linewidth, height = 5.5cm]{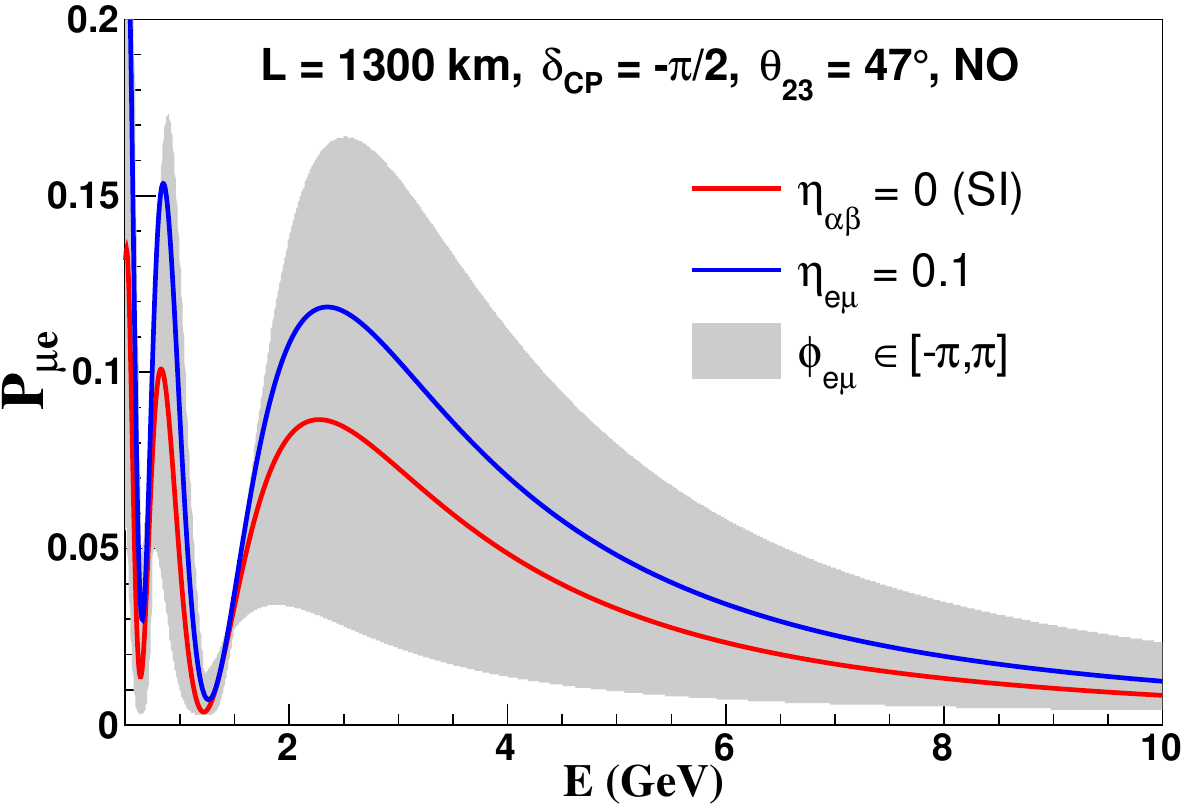}
    \includegraphics[width=0.328\linewidth, height = 5.5cm]{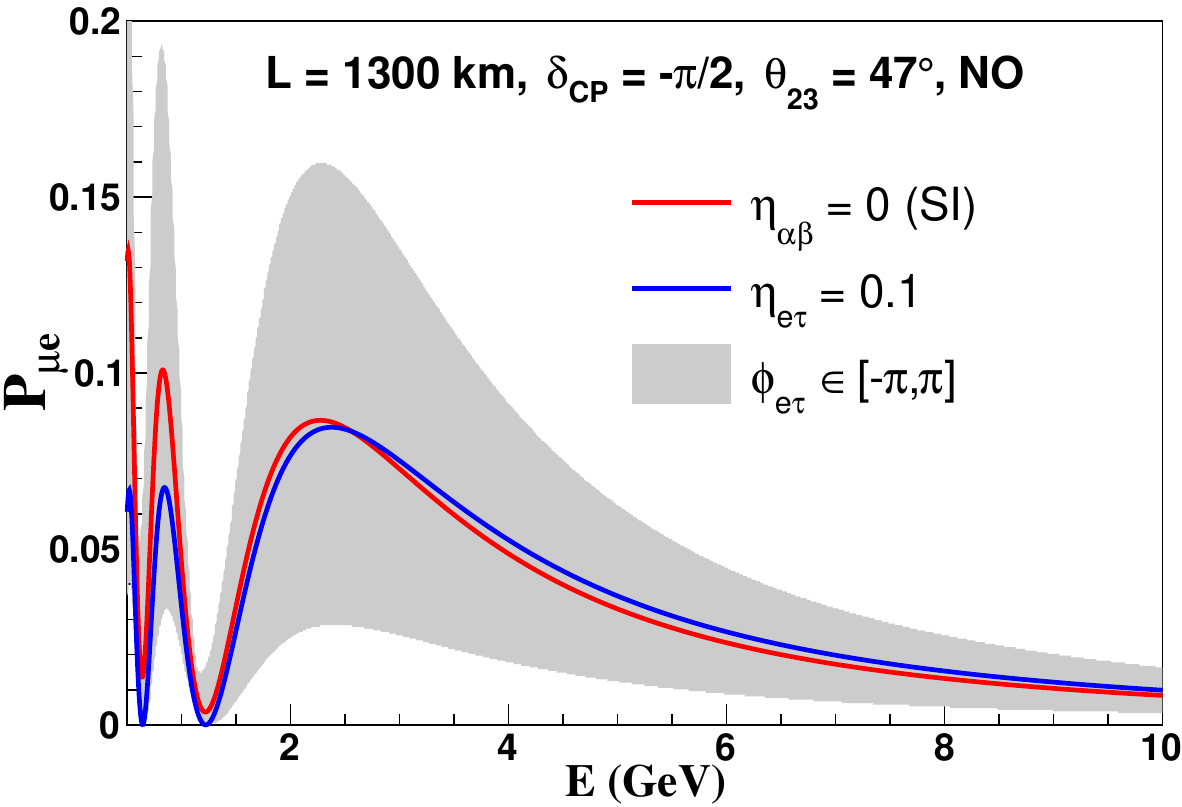}
    \includegraphics[width=0.328\linewidth, height = 5.5cm]{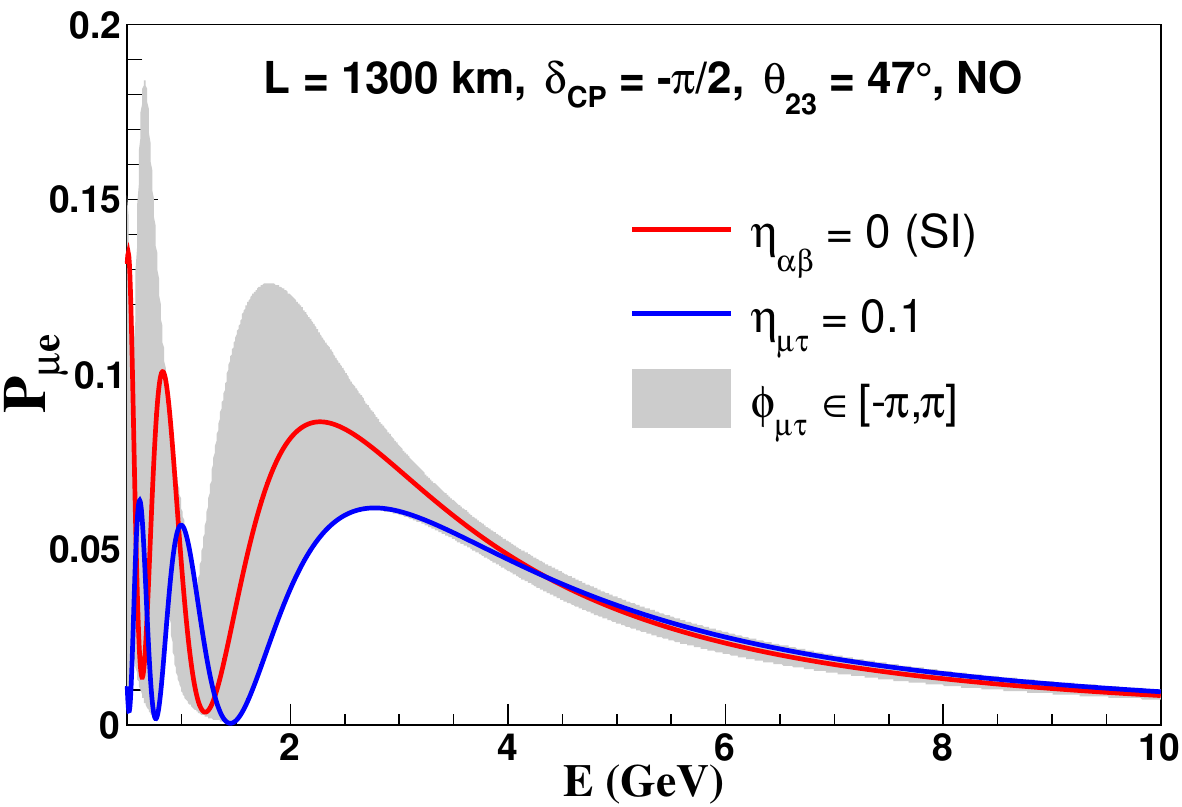}
    \caption{Impact on $P_{\mu e}$ due to the presence of off-diagonal elements $\eta_{e\mu}$, $\eta_{e\tau}$ and $\eta_{\mu\tau}$ for varying neutrino energy from 0.5-10 GeV for DUNE experiment i.e. a baseline of 1300km. The red and blue solid lines represent the SI and the scalar NSI case respectively. The shaded grey region showcases the variation due to the presence of scalar NSI $\phi_{\alpha\beta}$ phase.}
    \label{fig:Pmue}
\end{figure}

In figure \ref{fig:Pmue}, we have explored the impact of off-diagonal scalar NSI parameters ($|\eta_{e\mu}|$, $|\eta_{e\tau}|$, $|\eta_{\mu\tau}|$) and their associated phases ($\phi_{e\mu}$, $\phi_{e\tau}$, $\phi_{\mu\tau}$) on the appearance channel ($P_{\mu e}$). We study for the DUNE experiment which is with a baseline of 1300km. Here, we have considered only a single non-zero parameter at a time. We have listed all the values of neutrino oscillation parameters used for the simulation in table \ref{tab:param}. We have plotted $P_{\mu e}$ as a function of the neutrino energy which is varied in the range $0.5-10$ GeV. The probabilities are numerically calculated using the modified GLoBES framework. The red and blue solid lines represent the standard and the scalar NSI case ($|\eta_{\alpha\beta}|$=0.1) respectively. The impact on $P_{\mu e}$ due to the presence of corresponding phases $\phi_{\alpha\beta}$ is shown by the shaded grey band, where $\phi_{\alpha\beta}$ is varied in $[-\pi,\pi]$. 
\begin{itemize}
    \item In presence of $\eta_{e\mu}$, we see an enhancement in the probability values for varying neutrino energies. The non-zero values of the $\phi_{e\mu}$ phase can significantly modify the probabilities, as shown by the grey-shaded region. Depending on the value of the phase $\phi_{e\mu}$, we see either enhancement or suppression.
    
    \item Only nominal changes in the probabilities can be seen in the presence of $\eta_{e\tau}$. Although, the phase $\phi_{e\tau}$ can considerably enhance or suppress the values of $P_{\mu e}$. 
    
    \item In presence of $\eta_{\mu\tau}$, we observe  a noticeable suppression in the probability for $E<4$ GeV. In contrast to the presence of $\phi_{e\mu}$ and $\phi_{e\tau}$, the phase $\phi_{\mu\tau}$ results in a narrower band, indicating a smaller effect.
\end{itemize}

\begin{figure}[!b]
	\centering
    \includegraphics[width=0.328\linewidth, height = 5.5cm]{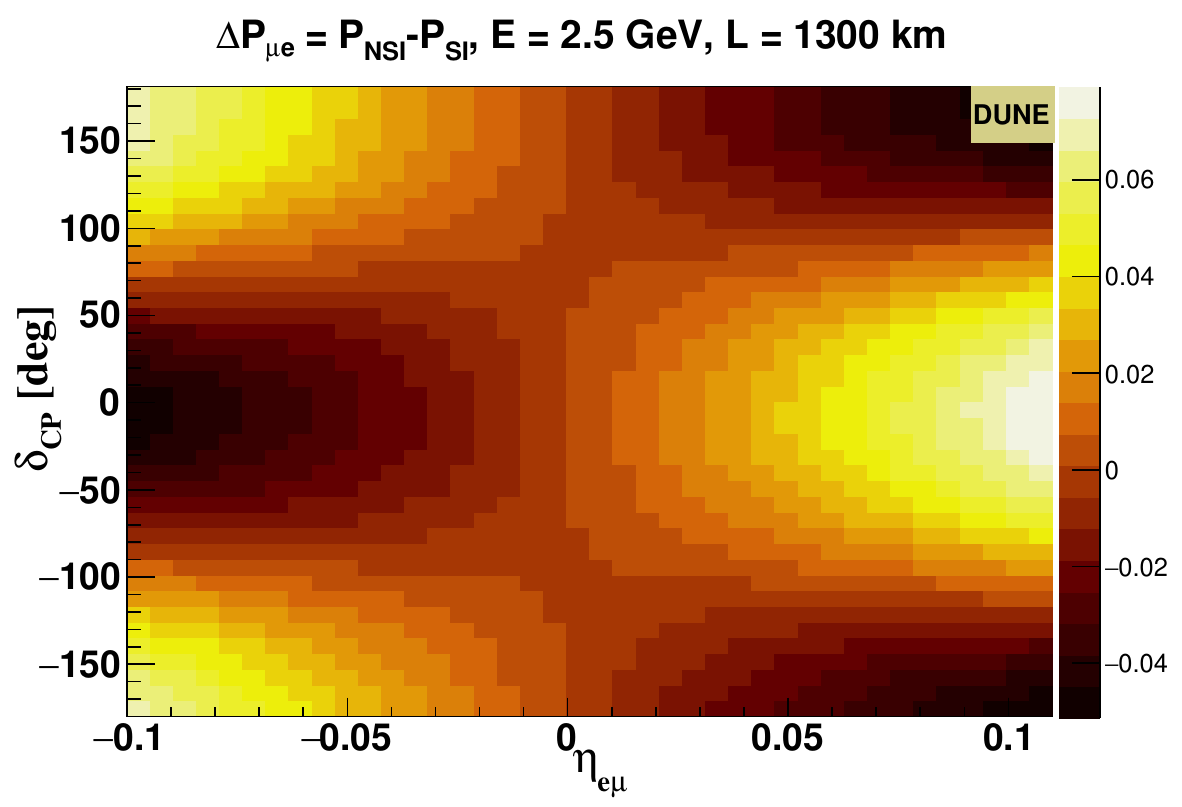}
    \includegraphics[width=0.328\linewidth, height = 5.5cm]{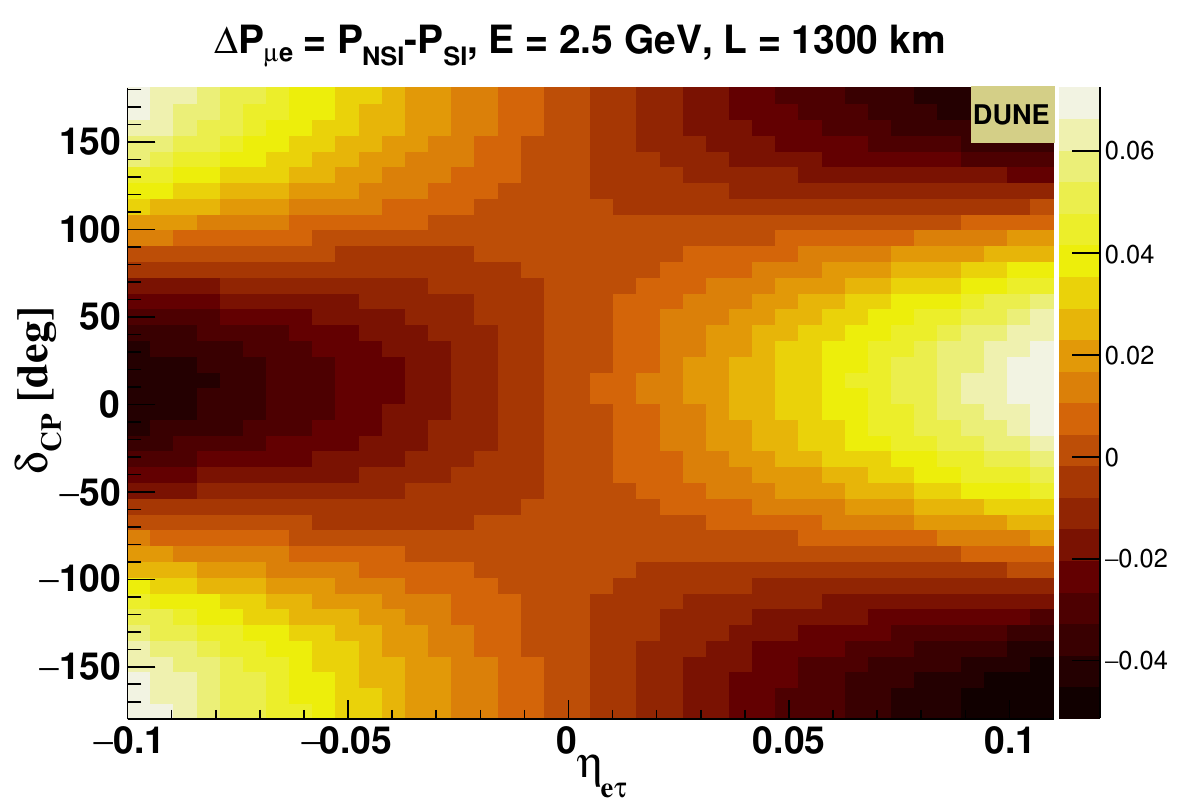}
    \includegraphics[width=0.328\linewidth, height = 5.5cm]{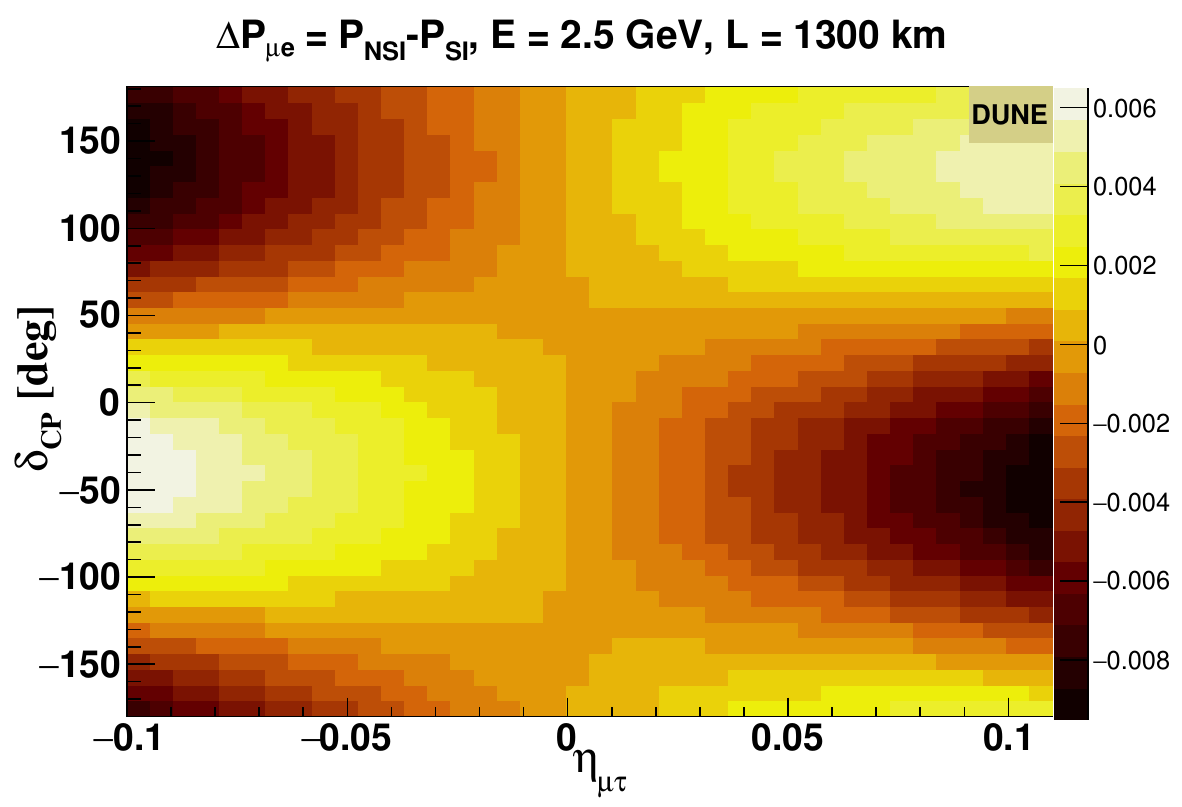}
    \caption{Plot of $\rm \Delta P_{\mu e} =\rm P_{\mu e}^{NSI}-P_{\mu e}^{SI}$ in the ($\eta_{\alpha\beta}$-$\delta_{CP}$) parameter space with neutrino energy fixed at the first oscillation maximum (i.e. 2.5 GeV) for DUNE with a baseline of 1300km. The vertical axes correspond to $\delta_{CP}$ and the horizontal axes correspond to $\eta_{\alpha\beta}$. The left, middle and right panels represent the case of $\eta_{e\mu}$, $\eta_{e\tau}$ and $\eta_{\mu\tau}$ respectively.}
    \label{fig:2d_delPmue}
\end{figure}

In order to quantitatively explore the impact of scalar NSI on $P_{\mu e}$ over the $\delta_{CP}$ parameter space, we define a quantity $\Delta P_{\mu e}$ as,

\begin{equation}
  \rm \Delta P_{\mu e} =\rm P_{\mu e}^{NSI}-P_{\mu e}^{SI}.  
\end{equation}

\noindent Here, $P_{\mu e}^{NSI}$ is the numerically calculated appearance probability in the presence of scalar NSI and $P_{\mu e}^{SI}$ is the appearance probability in the absence of scalar NSI. In figure \ref{fig:2d_delPmue}, we study the impact of ($\eta_{e\mu}$, $\eta_{e\tau}$, $\eta_{\mu\tau}$) on $P_{\mu e}$ in the complete $\delta_{CP}$ parameter space. We examine the effects of scalar NSI on $P_{\mu e}$ for the whole $\delta_{CP}$ parameter space to look for potential regions of degeneracies at the probability level. We have plotted $\Delta P_{\mu e}$ with $\delta_{CP}$ varied $\in[-\pi,\pi]$ and $\eta_{\alpha\beta}$ in [-0.1,0.1]. Note that the other mixing parameters have been set to the values described in the table \ref{tab:param}. The left, middle and right panels represent the $\eta_{e\mu}$, $\eta_{e\tau}$ and $\eta_{\mu\tau}$ cases, respectively.

\begin{itemize}
    \item In presence of $\eta_{e\mu}$ (left panel), we observe different regions in the $(\delta_{CP}$-$\eta_{e\mu})$ plane having similar impact of scalar NSI on $\Delta P_{\mu e}$. The degeneracies observed at the probability level indicate a possible degeneracy in the measurement of $\delta_{CP}$.
    \item We observe a similar pattern of degenerate regions in the presence of $\eta_{e\tau}$ (middle panel). The effects of $\eta_{e\mu}$ and $\eta_{e\tau}$ appear symmetric about $\delta_{CP}=0^{\circ}$.
    \item For $\eta_{\mu\tau}$ (right panel), we observe an opposite impact on $\Delta P_{\mu e}$ in comparison to $\eta_{e\mu}$ and $\eta_{e\tau}$. Additionally, the magnitude of the impact of $\eta_{\mu\tau}$ is also relatively less. 
\end{itemize}

\section{CP-measurement sensitivities in presence of $\eta_{\alpha\beta}$} \label{sec:CPmeasurement}

We observe that the presence of off-diagonal scalar NSI elements along with their associated phases can significantly modify the oscillation probabilities. Several regions of degeneracy at the probability level are also observed. This may affect the $\delta_{CP}$ measurement sensitivities at the upcoming long-baseline experiment DUNE. In section \ref{sec:dcp_precision}, we first explore the $\delta_{CP}$-constraining capability at DUNE in the presence of scalar NSI elements $\eta_{\alpha\beta}$. We also study the impact of scalar NSI on the CPV sensitivities at DUNE in section \ref{sec:CPVsens}.

\subsection{$\delta_{CP}$-Constraining capability} \label{sec:dcp_precision}
We explore the $\delta_{CP}$ constraining capability in the presence of off-diagonal scalar NSI elements ($\eta_{e\mu}$, $\eta_{e\tau}$, $\eta_{\mu\tau}$) at DUNE. We investigate the impact of $|\eta_{\alpha\beta}|$ and corresponding phases $\phi_{\alpha\beta}$ on the $\delta_{CP}$-measurement. 
The confidence levels (CL) of 3$\sigma$, 2$\sigma$ and 1$\sigma$ are represented by the black, red and blue solid lines, respectively. We have assumed normal mass ordering as the true ordering. All the oscillation parameters are fixed at the values shown in table \ref{tab:param}. 

In figure \ref{fig:eta_dcp}, we have a two-dimensional plane of ($\delta_{CP}$-$|\eta_{\alpha\beta}|$) with varying $\delta_{CP}$ along the horizontal axes and $|\eta_{\alpha\beta}|$ along the vertical axes. We have varied the test values of $\delta_{CP}$ in the complete parameter space $[-\pi,\pi]$ and $|\eta_{\alpha\beta}|$ in the range [0,0.1].  We have also marginalized over the oscillation parameters $\theta_{23}$, $\Delta m_{31}^{2}$ and corresponding phases $\phi_{\alpha\beta}$. We have considered two cases where, in the first case, the true value of ($\delta_{CP}^{true}$,$|\eta_{\alpha\beta}^{true}|$) is fixed at ($-90^{\circ}$,0.05) and the corresponding CL's are represented by the solid lines. In the second case, the true value is fixed at ($0^{\circ}$,0.05) and the CL's are represented by the dashed lines. The black star signifies the true values. The observations are listed below as:

\begin{figure}[!h]
	\centering
    \includegraphics[width=0.325\linewidth, height = 5.5cm]{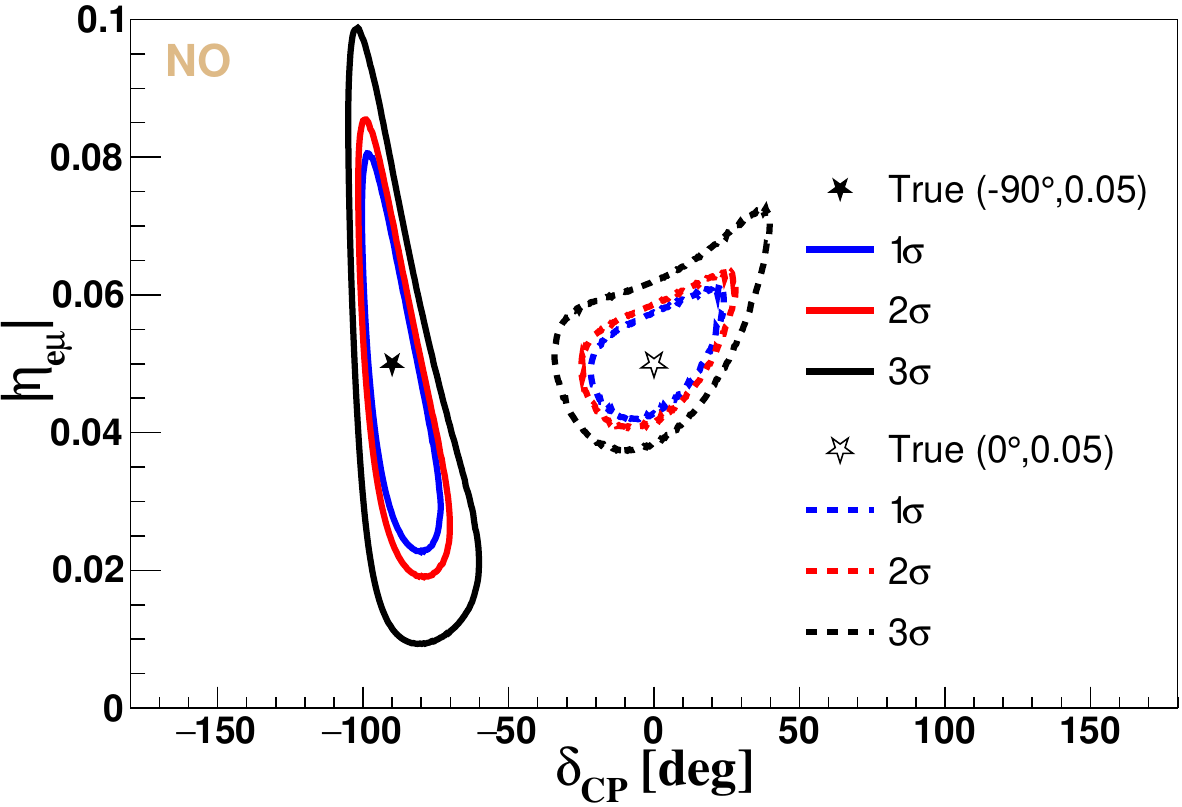}
    \includegraphics[width=0.325\linewidth, height = 5.5cm]{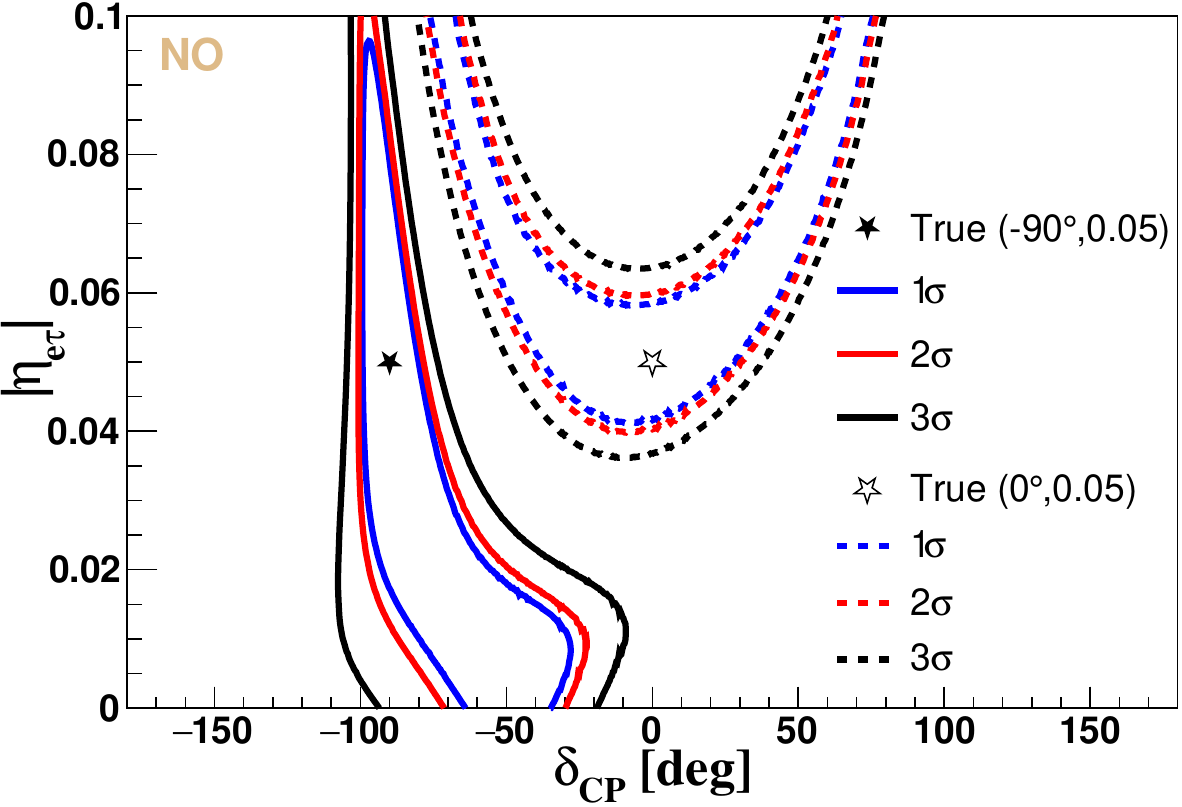}
    \includegraphics[width=0.325\linewidth, height = 5.5cm]{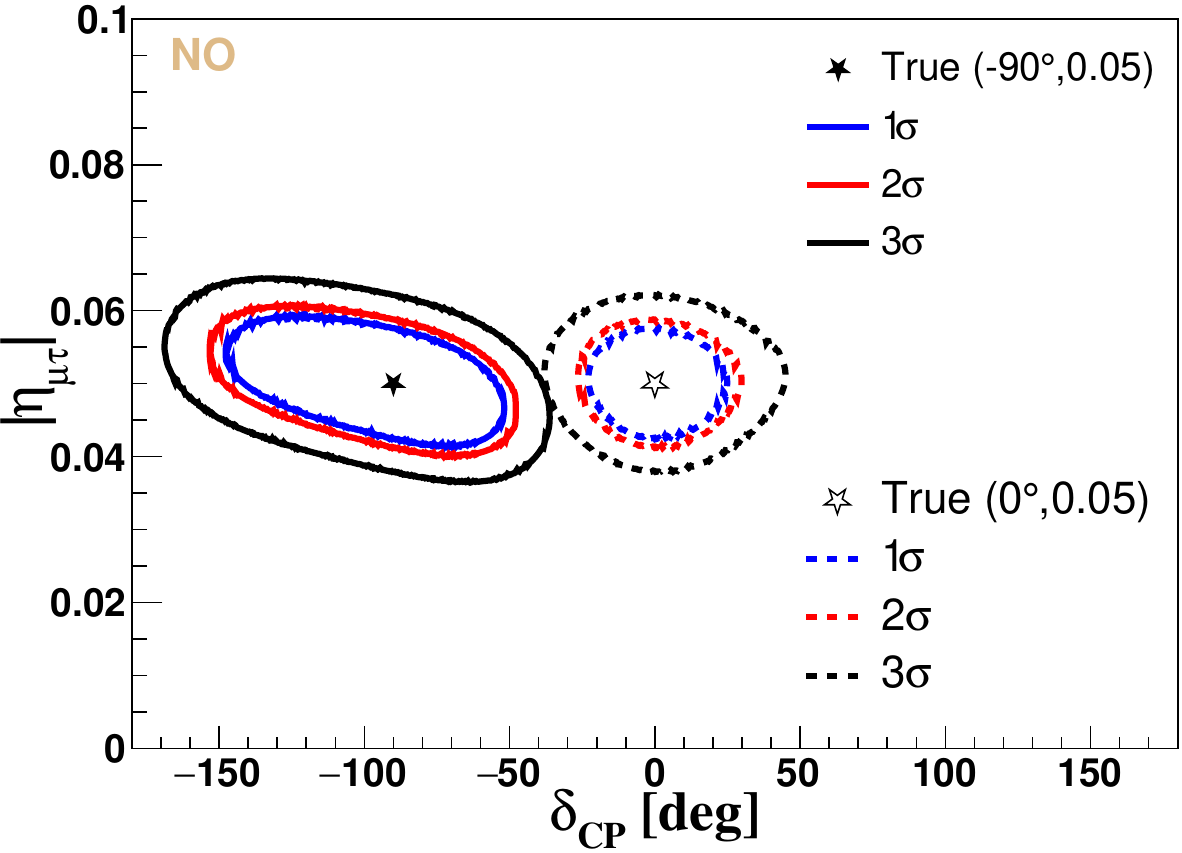}
    \caption{Sensitivity of DUNE to $|\eta_{e\mu}|$ (left-panel), $|\eta_{e\tau}|$ (middle-panel), $|\eta_{\mu\tau}|$ (right-panel) and the leptonic CP-phase $\delta_{CP}$. The blue, red and black colored lines represent the 1$\sigma$, 2$\sigma$ and 3$\sigma$ CL respectively. However, the solid CL lines represent the case when ($\delta_{CP}^{true}$,$|\eta_{\alpha\beta}^{true}|$) is fixed at ($-90^{\circ}$,0.05) and the dashed CL lines represent the case when ($\delta_{CP}^{true}$,$|\eta_{\alpha\beta}^{true}|$) is fixed at ($0^{\circ}$,0.05). The black star represents the true values.}
    \label{fig:eta_dcp}
\end{figure}

\begin{itemize}
    \item In the left panel, with true values of ($\delta_{CP}^{true}$,$|\eta_{\alpha\beta}^{true}|$) fixed at $(-90^{\circ},0.05)$, we observe that $\delta_{CP}$ is well constrained in the range $\delta_{CP}\in[-110^{\circ},-60^{\circ}]$ in presence of $\eta_{e\mu}$ but spreads widely over the $|\eta_{e\mu}|$ values. The allowed region remains constrained in the negative half-plane of $\delta_{CP}$. For $\delta_{CP}^{true}=0^{\circ}$, we observe a relatively better constraint about $|\eta_{e\mu}|$ but the constraining capability along $\delta_{CP}$ reduces.

    \item In the middle panel, we observe a similar constraint on $\delta_{CP}$ restricted only in the negative plane within the range $\delta_{CP}\in[-110^{\circ},-10^{\circ}]$ for $\eta_{e\tau}$ with ($\delta_{CP}^{true}$,$|\eta_{\alpha\beta}^{true}|$) fixed at $(-90^{\circ},0.05)$. However, it spreads in the complete $\eta_{e\tau}$ range. For $\delta_{CP}^{true}=0^{\circ}$, we observe a weaker constraint on the $\delta_{CP}$ with range $\delta_{CP}\in[-80^{\circ},80^{\circ}]$.
    
    \item In the right panel, we see that $\delta_{CP}$ is much better constrained in the presence of $\eta_{\mu\tau}$ compared to $\eta_{e\mu}$ and $\eta_{e\tau}$ for both scenarios. However, the constraint is much better for the scenario $\delta_{CP}^{true}=0^{\circ}$ in comparison to $\delta_{CP}^{true}=-90^{\circ}$. 
\end{itemize}

\begin{figure}[!h]
	\centering
    \includegraphics[width=0.325\linewidth, height = 5.5cm]{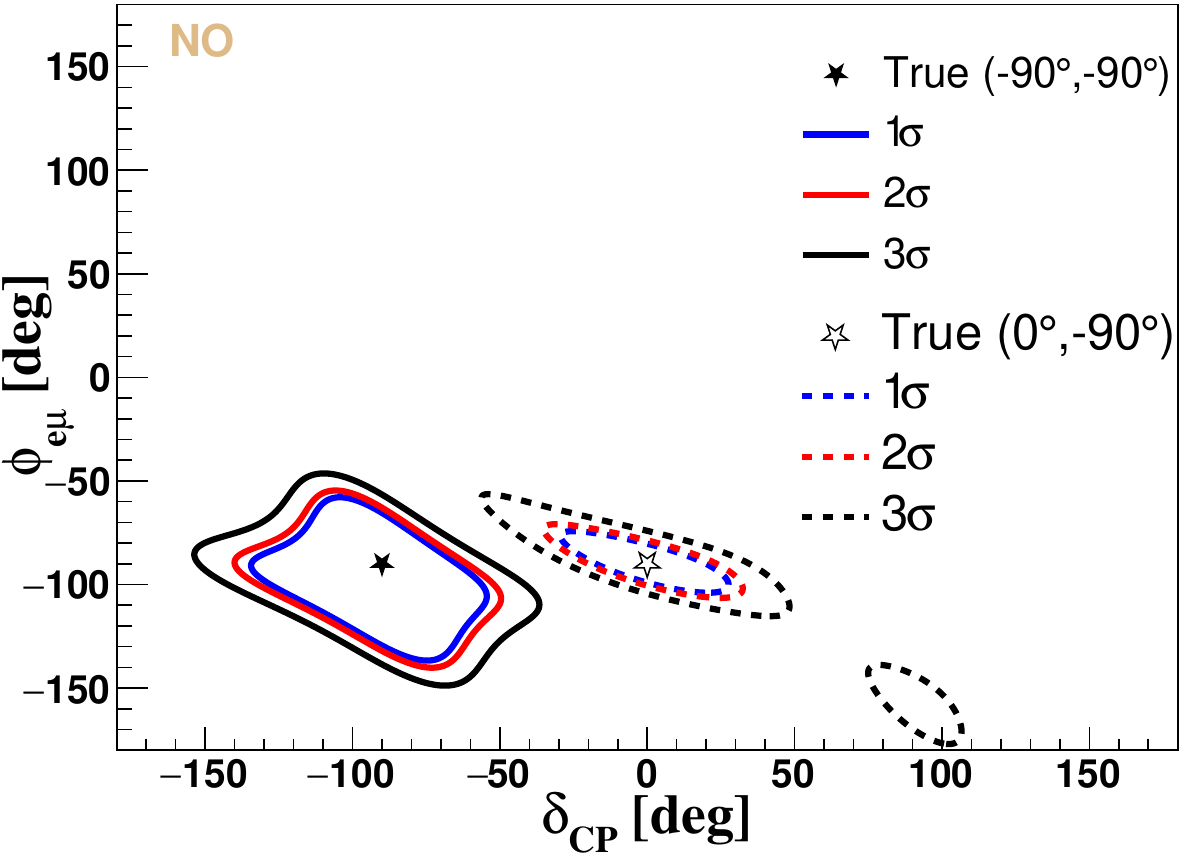}
    \includegraphics[width=0.325\linewidth, height = 5.5cm]{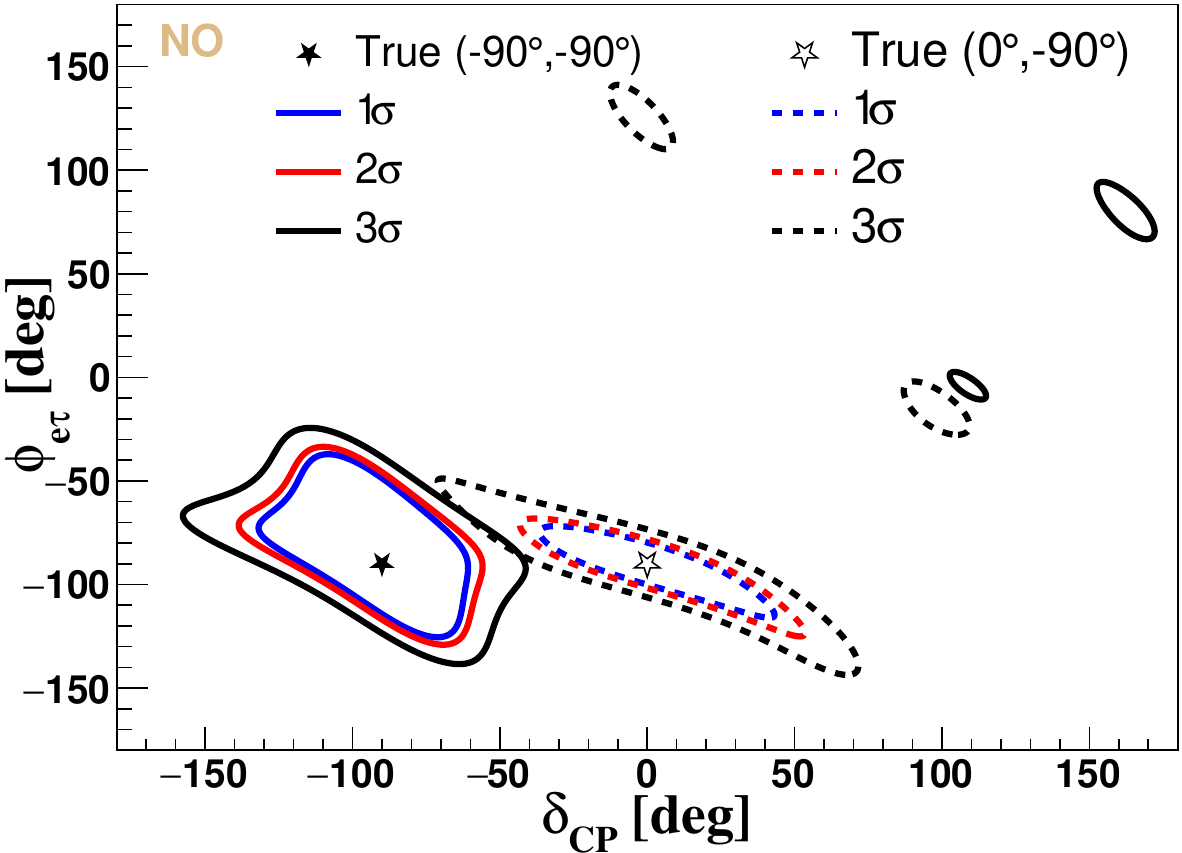}
    \includegraphics[width=0.325\linewidth, height = 5.5cm]{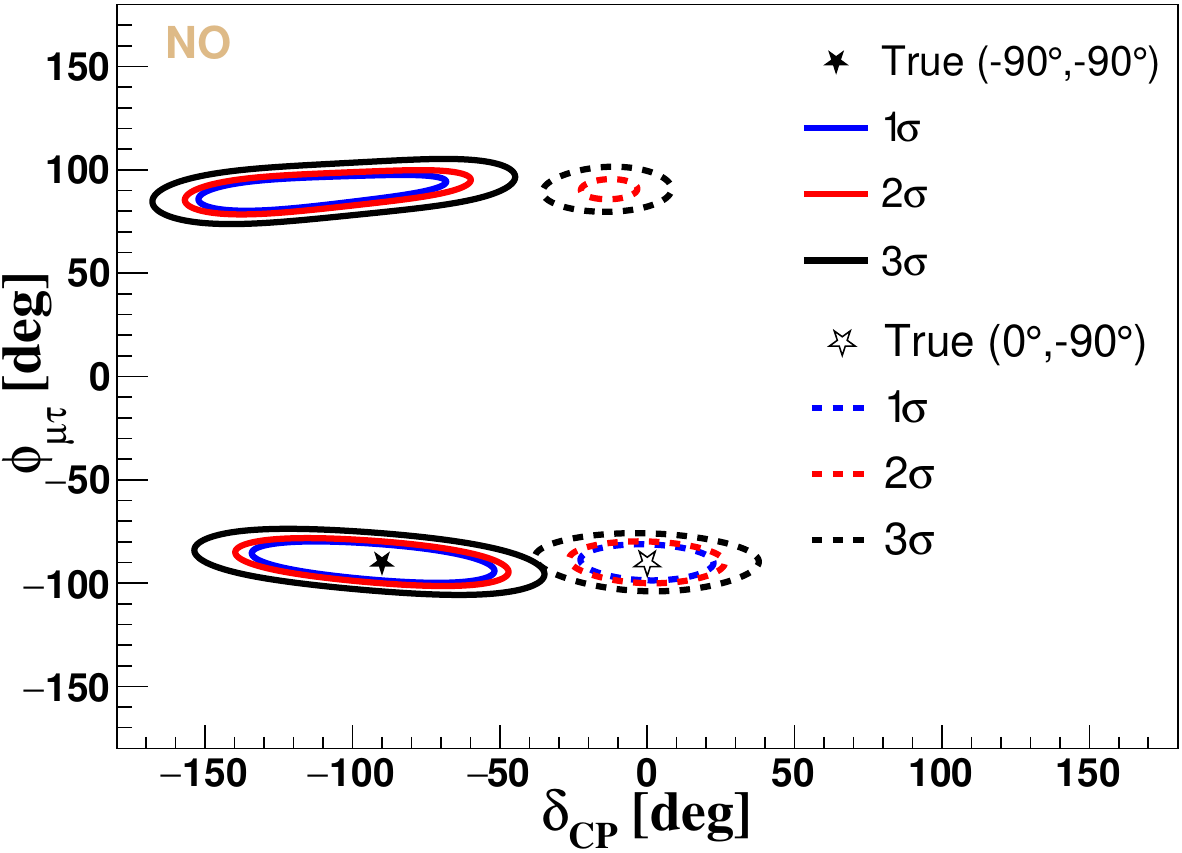}
    \caption{Sensitivity of DUNE to $\phi_{e\mu}$ (left-panel), $\phi_{e\tau}$ (middle-panel), $\phi_{\mu\tau}$ (right-panel) and  the leptonic CP-phase $\delta_{CP}$. The blue, red and black lines represent the 1$\sigma$, 2$\sigma$ and 3$\sigma$ CL respectively. In all the panels, the solid lines represent the case when ($\delta_{CP}^{true}$,$\phi_{\alpha\beta}^{true}$) is fixed at ($-90^{\circ}$,$-90^{\circ}$) and the dashed lines represent the case when ($\delta_{CP}^{true}$,$\phi_{\alpha\beta}^{true}$) is fixed at ($0^{\circ}$,$-90^{\circ}$). The black star represents the true values.}
    \label{fig:phi_dcp}
\end{figure}

\noindent In figure \ref{fig:phi_dcp}, we plot a two-dimensional plane spanning the ($\delta_{CP}$-$\phi_{\alpha\beta}$) parameter space, where the color scheme is the same as used in figure \ref{fig:eta_dcp}. Similar to the previous figure, we consider two different scenarios in this analysis. In the first scenario, we consider the true values of ($\delta_{CP}^{true}$,$\phi_{\alpha\beta}^{true}$) are fixed at ($-90^{\circ}$,$-90^{\circ}$) and the resulting CL's are shown by the solid lines. In the second scenario, the true value is fixed at ($0^{\circ}$,$-90^{\circ}$) and the CL's are shown by dashed lines. The black star signifies the true values. All the other oscillation parameters used for simulation are as described in table \ref{tab:param}. We have fixed $|\eta_{\alpha\beta}^{true}|=0.05$ for all the cases and varied $\phi_{\alpha\beta}$ in the complete parameter space.

\begin{itemize}
    \item In left panel with ($\delta_{CP}^{true}$,$\phi_{\alpha\beta}^{true}$) fixed at $(-90^{\circ},-90^{\circ})$, we see that $\delta_{CP}$ is well constrained in the negative half-plane within the range $\delta_{CP}\in[-160^{\circ},-40^{\circ}]$. For $\delta_{CP}^{true}=0^{\circ}$, we observe a much better constraint on $\delta_{CP}$ with $\delta_{CP}\in[-50^{\circ},50^{\circ}]$. However, we find a degenerate region in the positive $\delta_{CP}$ plane around $\delta_{CP}\sim90^{\circ}$ at $3\sigma$ CL.

    \item In presence of $\phi_{e\tau}$, the constraint on $\delta_{CP}$ is relatively weaker in comparison to $\phi_{e\mu}$ as seen in the middle panel. In this case, we observe multiple degenerate regions for both scenarios, i.e., $\delta_{CP}^{true}=-90^{\circ}$ and $0^{\circ}$.

    \item The presence of $\phi_{\mu\tau}$ leads to degenerate regions which can be prominently seen in the right panel. The degenerate regions in positive and negative $\phi_{\mu\tau}$ plane is observed for $\phi_{\mu\tau}\sim\pm90^{\circ}$. This can compromise DUNE's sensitivity towards CP-violation in the presence of scalar NSI. We make similar observations for both $\delta_{CP}^{true}$ scenarios. However, the constraining is significantly better for $\delta_{CP}^{true}=0^{\circ}$. 
\end{itemize}

We see that the presence of off-diagonal scalar NSI can significantly affect the $\delta_{CP}$ measurement. The non-zero phases $\phi_{\alpha\beta}$ can give rise to degeneracies and further hamper the CP-measurement sensitivities at DUNE.

\subsection{Impact on CP-violation sensitivities} \label{sec:CPVsens}
We explore the sensitivity of DUNE towards measuring the leptonic CP-violating phase $\delta_{CP}$ in the presence of off-diagonal scalar NSI parameters ($\eta_{e\mu}$, $\eta_{e\tau}$, $\eta_{\mu\tau}$). CPV sensitivity quantifies how well an experiment can distinguish between CP-conserving ($\delta_{CP}=[0,\pm \pi$]) and CP-violating values of $\delta_{CP}$. For the analysis, we can define the sensitivity for CPV as
\begin{equation}
{\Delta \chi}^{2}_{\rm CPV}~(\delta_{\rm CP}) = {\rm min}~\left[\chi^2~(\delta_{CP},\delta^\text{test}_{CP}=0),~\chi^2 (\delta_{CP},\delta^\text{test}_{CP}=\pm \pi)\right ].
\end{equation}

\begin{figure}[!h]
    \centering
    \includegraphics[width=0.328\linewidth, height = 5.5cm]{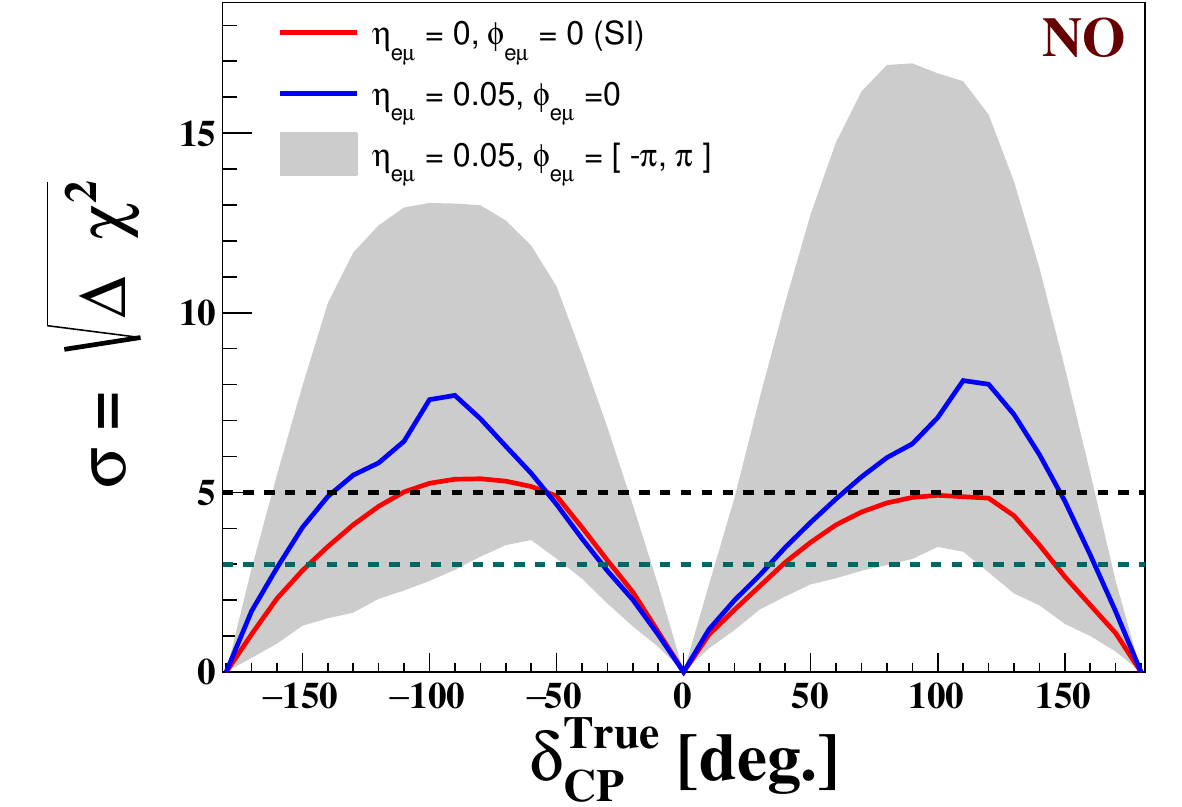}
    \includegraphics[width=0.328\linewidth, height = 5.5cm]{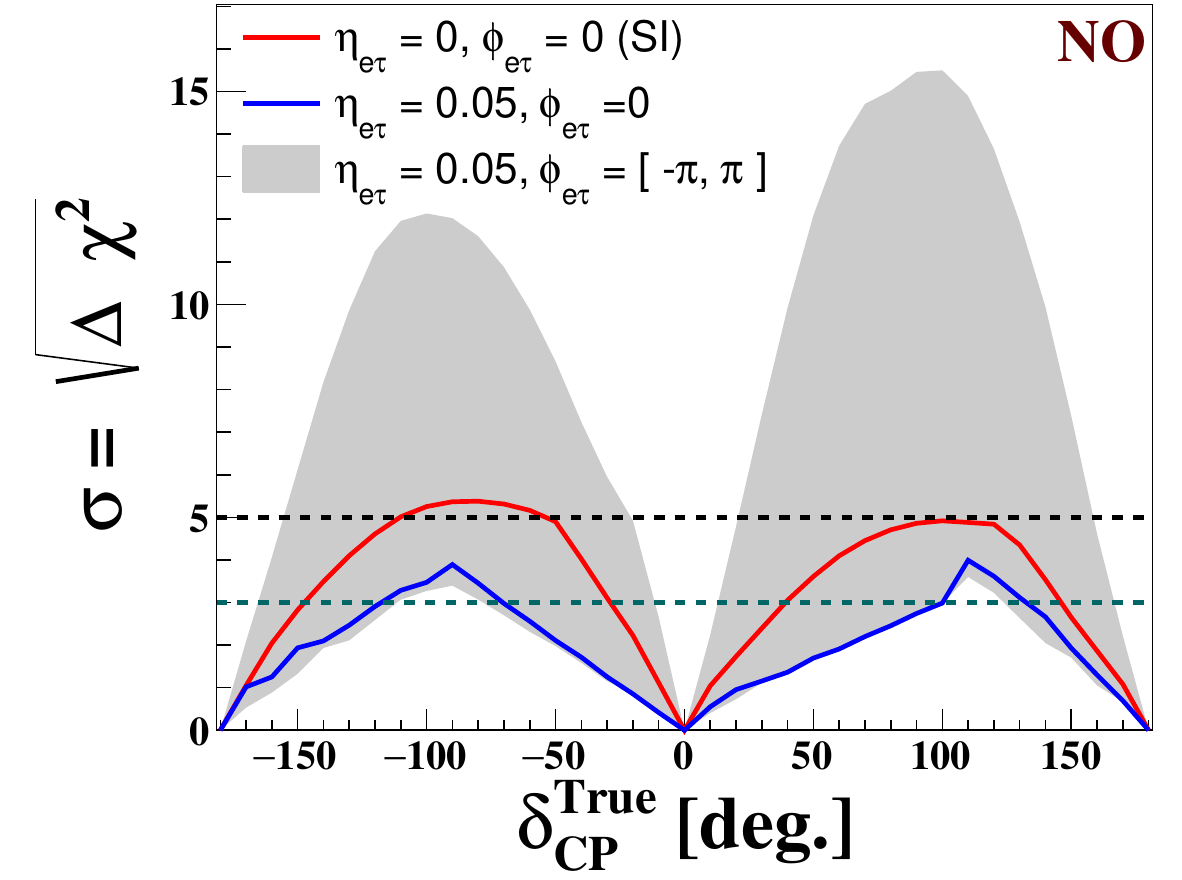}
    \includegraphics[width=0.328\linewidth, height = 5.5cm]{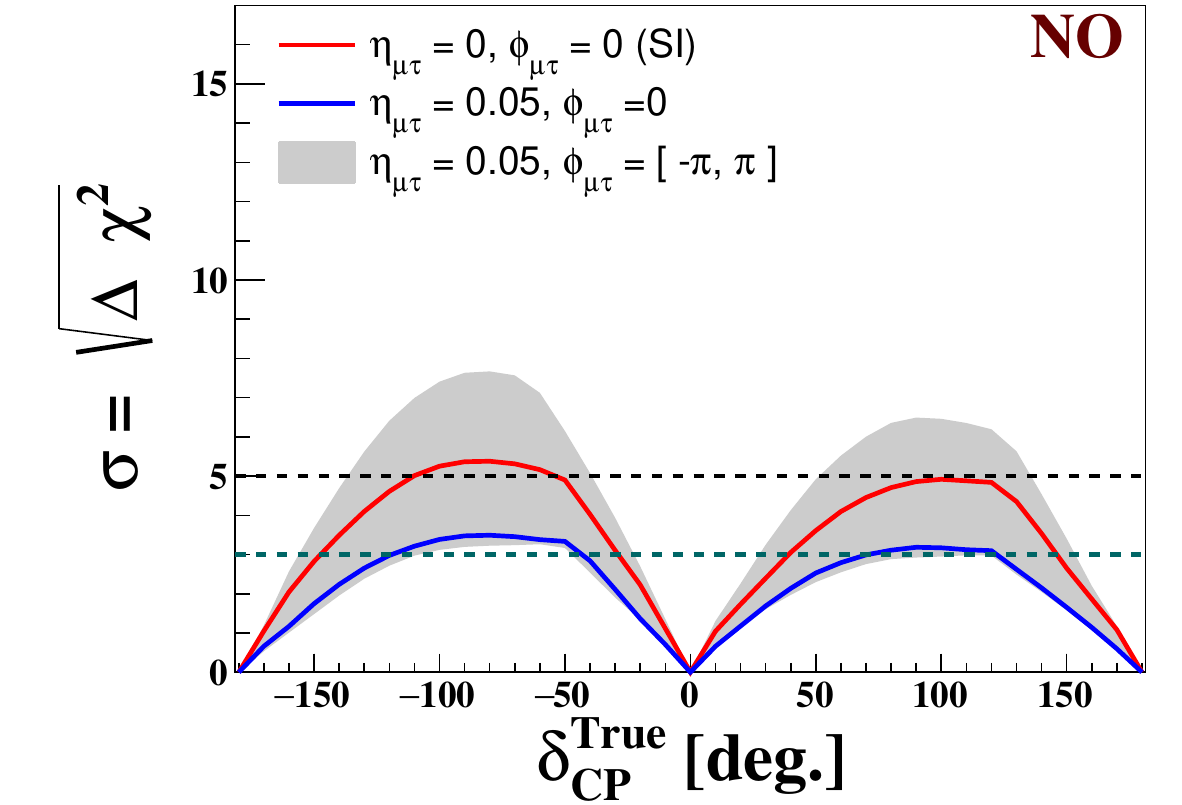}
    \caption{CPV sensitivity of DUNE in the presence of off-diagonal scalar NSI parameters $|\eta_{e\mu}|$ (left-panel), $|\eta_{e\tau}|$ (middle-panel), $|\eta_{\mu\tau}|$ (right-panel). The red and blue solid lines correspond to SI and $|\eta_{\alpha\beta}|=0.05$. The grey band signifies the variation of $\phi_{\alpha\beta}$ in the range $[-\pi,\pi]$. The dashed green and black lines represent $3\sigma$ and $5\sigma$ CL, respectively.}
    \label{fig:CPVsens}
\end{figure}

We obtain the minimum of $\Delta\chi^{2}_{CPV}$ for all the $\delta_{CP}$ values to measure CP-violation in the complete $\delta_{CP}$ parameter space. We assume the normal mass ordering as the true ordering and a higher $\theta_{23}$ octant as the true octant. We define $\sigma=\sqrt{{\Delta \chi}^{2}}$ as the CPV significance, where 3$\sigma$ (5$\sigma$) CL are represented by the horizontal lines at $\sqrt{{\Delta \chi}^{2}}$ = 9 (25) respectively. We have marginalized the standard oscillation parameters in their 3$\sigma$ allowed range as shown in table \ref{tab:param}. We have also additionally marginalized over the scalar NSI parameters $|\eta_{\alpha \beta}| \in [0,0.1]$. In all panels, the red solid line corresponds to the SI case, whereas the blue solid line represents the $|\eta_{\alpha\beta}|=0.05$ case. The green (black) dashed line representing 3$\sigma$ (5$\sigma$) CL is used as a reference. In figure \ref{fig:CPVsens}, we have plotted the significance, i.e., $\sigma=\sqrt{\Delta\chi^{2}}$ along the vertical axes and varied the $\delta_{CP}^{True}$ along the horizontal axes. In each panel, we have additionally varied the corresponding phase $\phi_{\alpha\beta}$ in the range $[-\pi,\pi]$. The impact on the CPV sensitivities for varying $\phi_{\alpha\beta}$ in the range $\in [-\pi,\pi]$ is represented by the shaded grey band. 
 
\begin{itemize}
    \item In presence of $\eta_{e\mu}$ (left panel), we observe an enhancement in the sensitivities. We also note that the presence of $\phi_{e\mu}$ can significantly alter the CPV sensitivities depending on its value.

    \item The presence of $\eta_{e\tau}$ (middle panel) deteriorates the sensitivities in comparison to the SI case for the complete $\delta_{CP}$ space. And the presence of the corresponding phase $\phi_{e\tau}$ can enhance/suppress the sensitivities as also seen for $\phi_{e\mu}$.
    
    \item Similar to $\eta_{e\mu}$, we observe that the presence of $\eta_{\mu\tau}$ (right panel) also suppresses the sensitivities in the complete $\delta_{CP}$ parameter space. Depending on the value of $\phi_{\mu\tau}$, we observe enhancement/suppression of the sensitivities. We also note that the variation of sensitivities due to $\phi_{\mu\tau}$ is relatively small compared to $\phi_{e\mu}$ and $\phi_{e\tau}$.
\end{itemize}

\begin{figure}[!h]
    \centering
    \includegraphics[width=0.328\linewidth, height = 5.5cm]{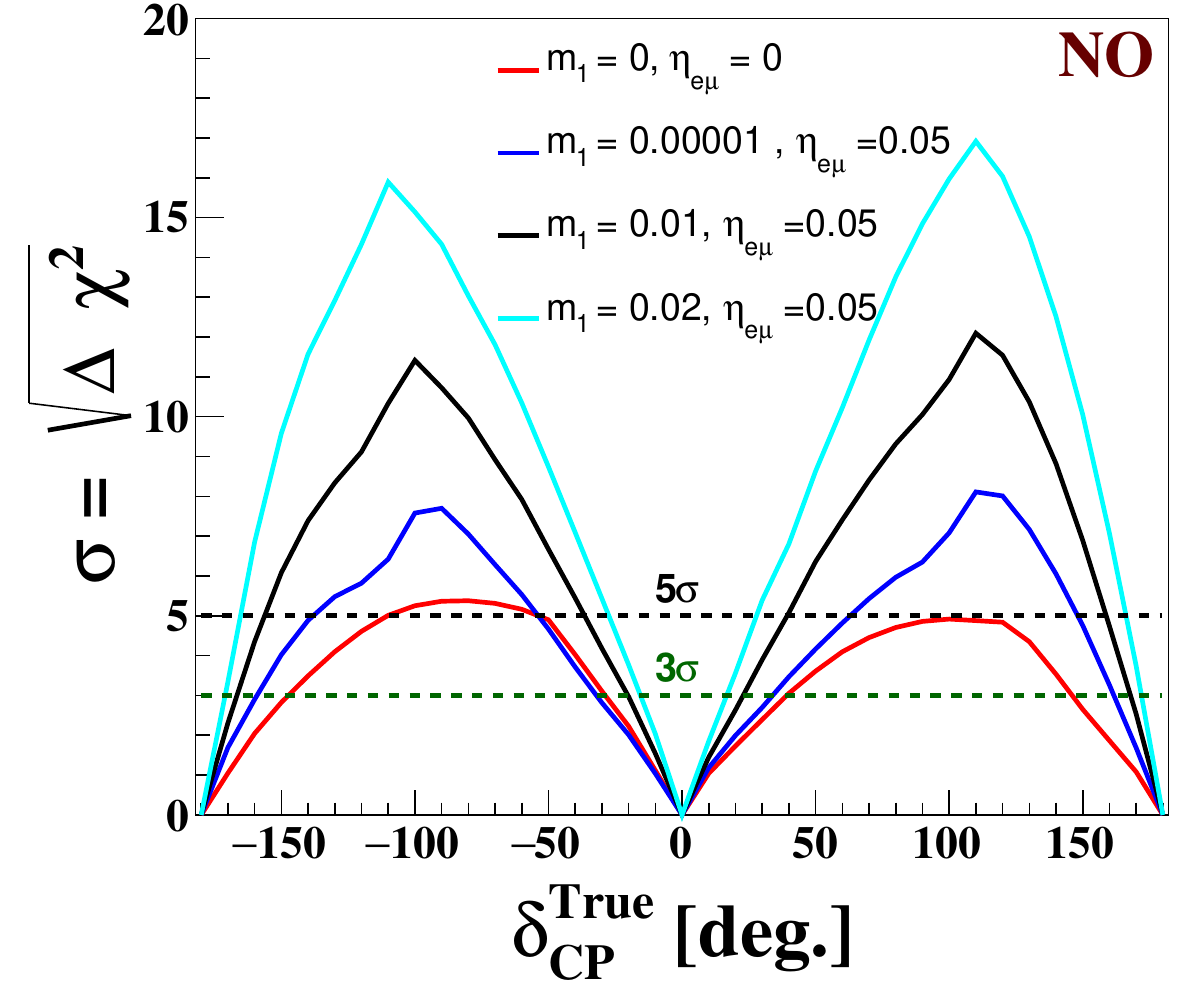}
    \includegraphics[width=0.328\linewidth, height = 5.5cm]{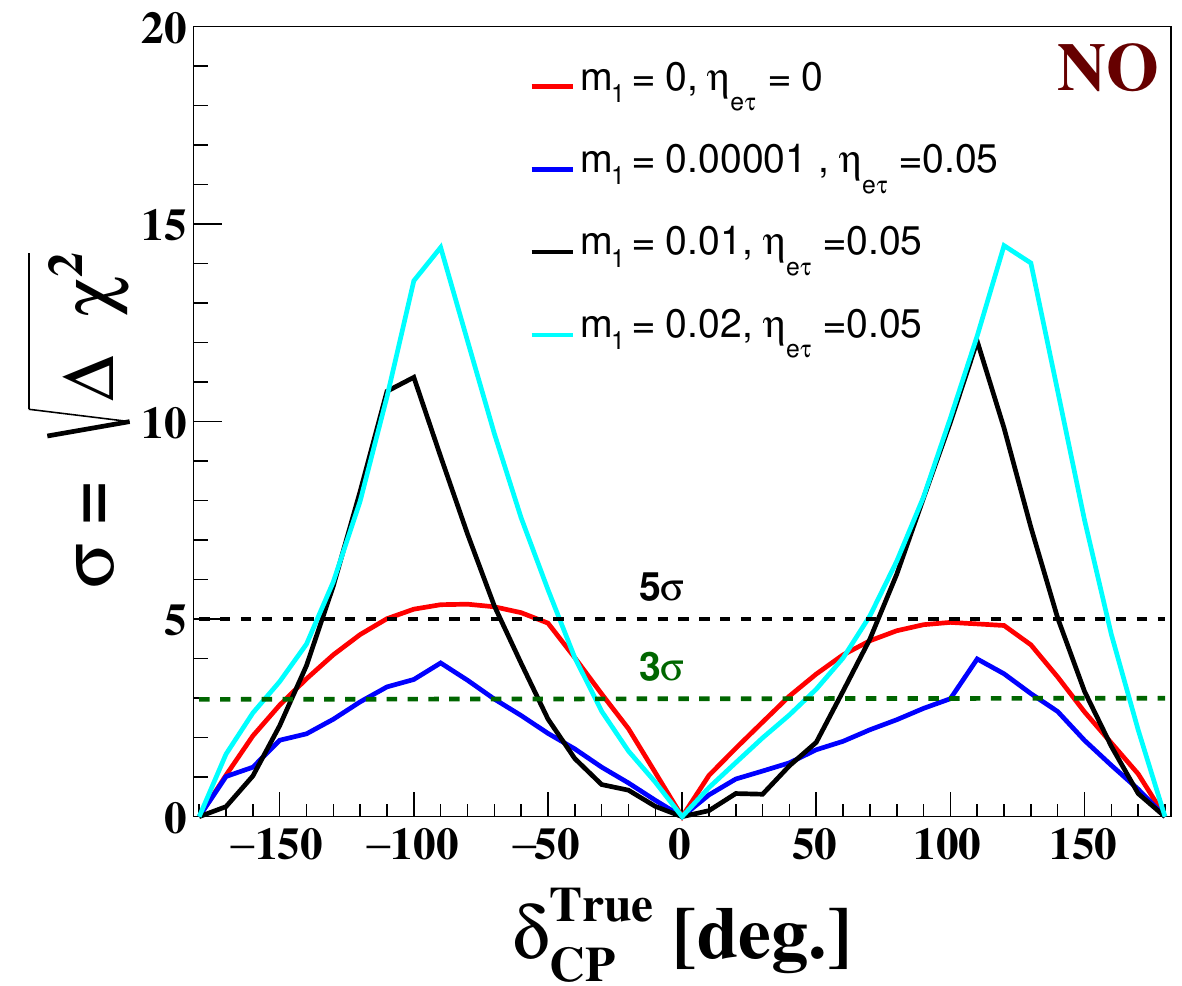}
    \includegraphics[width=0.328\linewidth, height = 5.5cm]{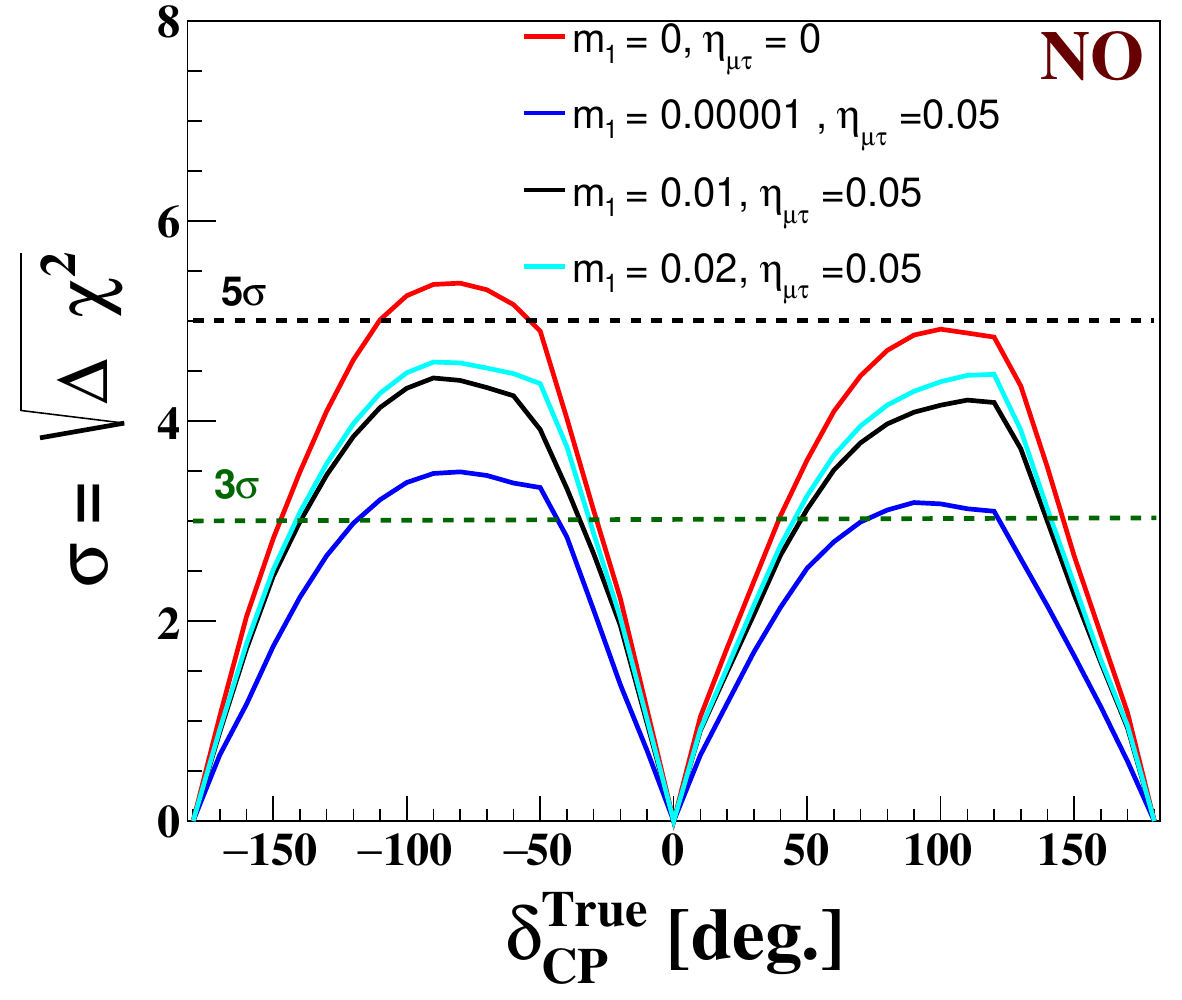}
    \caption{Impact of different neutrino mass scale on CPV sensitivity at DUNE in the presence of off-diagonal scalar NSI parameters $|\eta_{e\mu}|$ (left-panel), $|\eta_{e\tau}|$ (middle-panel), $|\eta_{\mu\tau}|$ (right-panel). The red solid line represents the SI case. The other coloured lines correspond to a different value of the lightest $\nu$-mass and $|\eta_{\alpha\beta}|=0.05$. The dashed green (black) line represents the $3\sigma$ ($5\sigma$) CL.}
    \label{fig:CPVmassImpact}
\end{figure}

In figure \ref{fig:CPVmassImpact}, we have explored the impact of different neutrino mass scale on the CPV sensitivities at DUNE. We have plotted for the scalar NSI parameters $|\eta_{e\mu}|$, $|\eta_{e\tau}|$ and $|\eta_{\mu\tau}|$ on the left, middle, and right panels, respectively. The values of the oscillation parameters are the same as those used in figure \ref{fig:CPVsens}. The red solid line represents the SI case. The blue, black and cyan solid lines correspond to the different choices of the lightest neutrino mass, i.e., $m_{1}=10^{-5}, 0.01, 0.02$ eV, respectively. For the above cases, the scalar NSI parameter is fixed at $\eta_{\alpha\beta}=0.05$.
\begin{itemize}
    \item We observe that the increase in the value of the lightest neutrino mass enhances the overall CPV sensitivities in the presence of parameters $\eta_{e\mu}$ and $\eta_{e\tau}$. 
    
    \item As shown in figure \ref{fig:CPVsens}, the presence of $\eta_{\mu\tau}$ (right-panel) can suppress the sensitivities in the complete $\delta_{CP}$ space. However, increasing the neutrino mass scale tends to restore the sensitivities to the SI case.
\end{itemize}

\section{Sensitivity to off-diagonal scalar NSI elements}\label{sec:eta_bound}

\begin{figure}[!h]
	\centering
    \includegraphics[width=0.45\linewidth, height = 5.5cm]{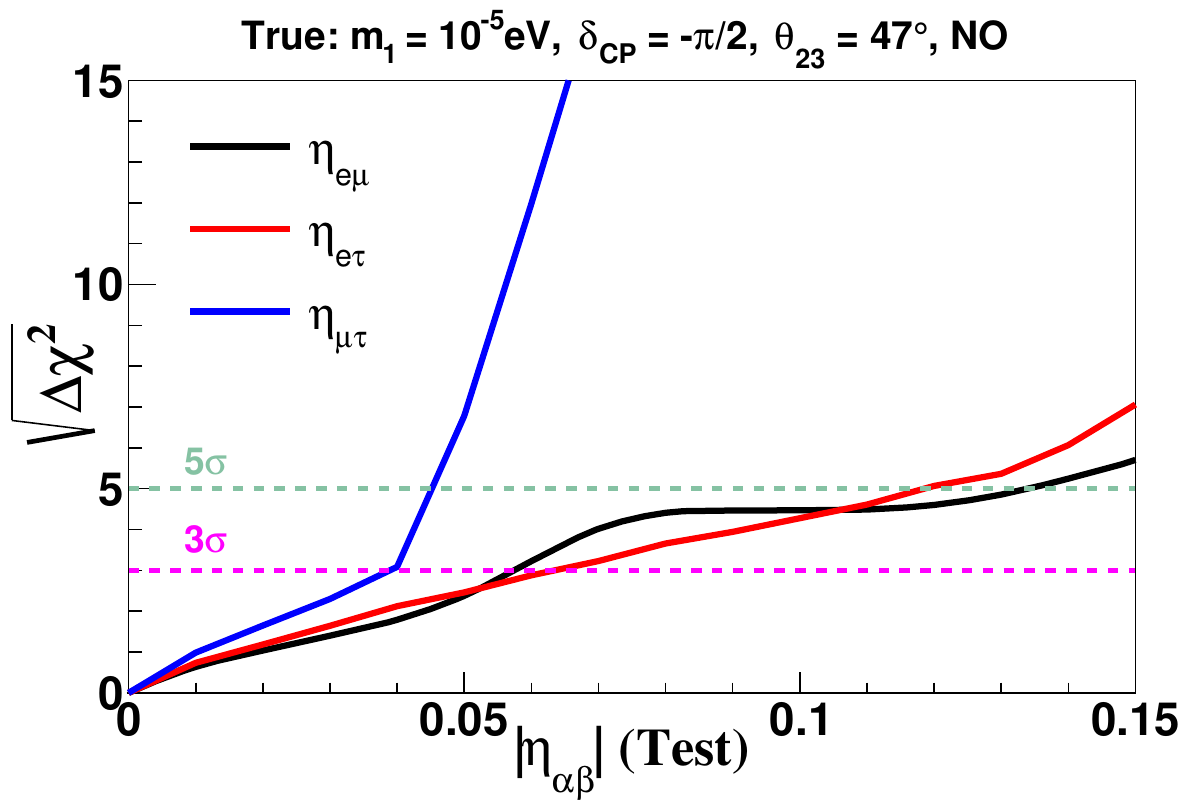}
    \includegraphics[width=0.45\linewidth, height = 5.5cm]{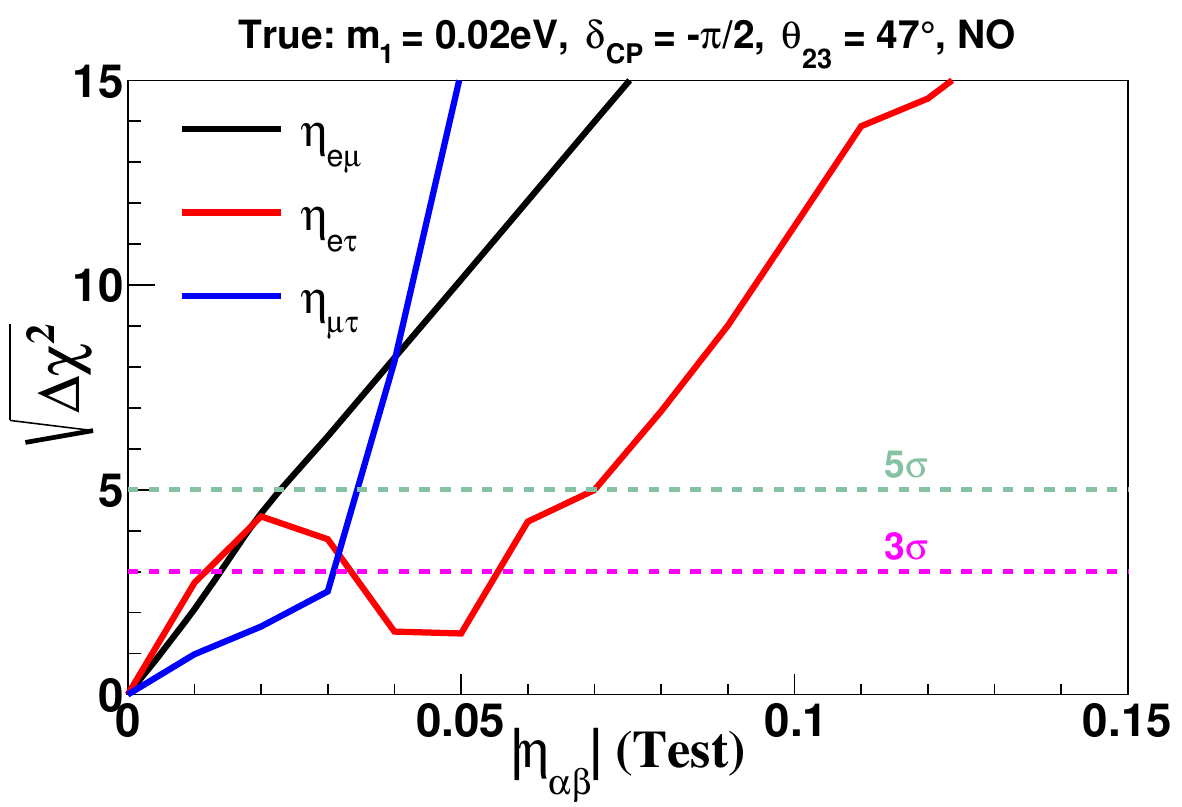}
    \caption{DUNE's sensitivity to the elements $|\eta_{e\mu}|$, $|\eta_{e\tau}|$ and $|\eta_{\mu\tau}|$ are shown by black, red and blue solid lines respectively. The dashed lines in magenta and green indicate the 3$\sigma$ and 5$\sigma$ CL. The left (right) panel corresponds to the lightest $\nu$-mass, $m_{1}=10^{-5}(0.02)$ eV.}
\label{fig:fix_chi2}
\end{figure}

We constrain the off-diagonal scalar NSI elements ($|\eta_{e\mu}|$, $|\eta_{e\tau}|$, $|\eta_{\mu\tau}|$) by considering the simulation of the DUNE experiment in the modified GLoBES framework. In the presence of scalar NSI, neutrino oscillation experiments can become sensitive to the neutrino mass scale. We have also probed the impact of different choices of the lightest neutrino mass on constraining the parameters. The oscillation parameters used for simulation are tabulated in table \ref{tab:param}. We have followed a similar marginalization for the standard oscillation parameters. In figure \ref{fig:fix_chi2}, we have constrained the off-diagonal scalar NSI elements $\eta_{e\mu}$, $\eta_{e\tau}$ and $\eta_{\mu\tau}$ which are represented by the black, red and blue solid lines, respectively. We have plotted the $\Delta \chi^{2}$ profiles about the vertical axes after marginalizing over the mixing parameters $\theta_{23}$, $\Delta m_{31}^{2}$. We have varied $|\eta_{\alpha\beta}|$ along the horizontal axes. We have considered normal ordering as the true ordering and the higher $\theta_{23}$ octant is considered as the true octant. The 3$\sigma$ and 5$\sigma$ CL are represented by the dashed lines in magenta and green, respectively. We have set the lightest neutrino mass at $m_{1}=10^{-5}$ eV in the left panel. On the other hand, we have it fixed at $m_{1}=0.02$ eV in the right panel.

\begin{itemize}
    \item We observe that DUNE is sensitive to the off-diagonal elements and can put strong constraints on the parameters. For $m_{1}=10^{-5}$ eV, we see that $\eta_{\mu\tau}$ is better constrained in comparison to $\eta_{e\mu}$ and $\eta_{e\tau}$ at $5\sigma$ CL. The constraining of $\eta_{e\mu}$ is comparable to that of $\eta_{e\tau}$ element.

    \item When the lightest neutrino mass is fixed at $m_{1}=0.02$ eV, we see that the constrain on $\eta_{\mu\tau}$ remains almost same as before. But the constraint on $\eta_{e\mu}$ significantly improves at $5\sigma$ CL. 

\end{itemize}

\begin{table}[!t]
    \centering
    \begin{tabular}{c|c|c}
        \hline
        Parameters & Bounds for $m_{1}=10^{-5}$ eV & Bounds for $m_{1}=0.02$ eV\\
        \hline
        $|\eta_{e\mu}|$ & 0.13 & 0.02\\
        $|\eta_{e\tau}|$ & 0.12 & 0.06\\
        $|\eta_{\mu\tau}|$ & 0.045 & 0.035\\
        \hline
    \end{tabular}
    \caption{Bounds on the parameters at $5\sigma$ CL obtained from the simulation of DUNE.}
    \label{tab:DUNE_cons_offdiag}
\end{table}

The bounds on the off-diagonal scalar NSI parameters at $5\sigma$ CL as obtained from the simulation of DUNE experiment are shown in table \ref{tab:DUNE_cons_offdiag}. Overall, the increase in the neutrino mass scale leads to a better constraining of the off-diagonal elements. In addition, the constraining is significantly better, particularly for the elements $\eta_{e\mu}$ and $\eta_{e\tau}$.

\section{Correlation among the $\eta_{\alpha\beta}$ parameters}\label{sec:corr}

\begin{figure}[!h]
    \centering
    \includegraphics[width=0.7\linewidth]{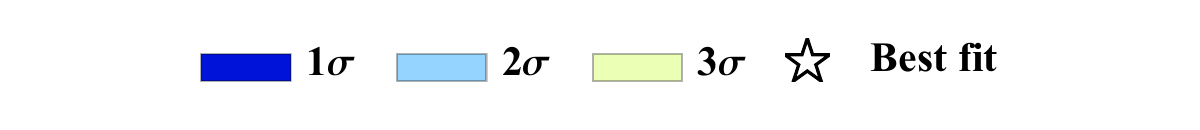}
    \includegraphics[width=0.328\linewidth, height=5cm]{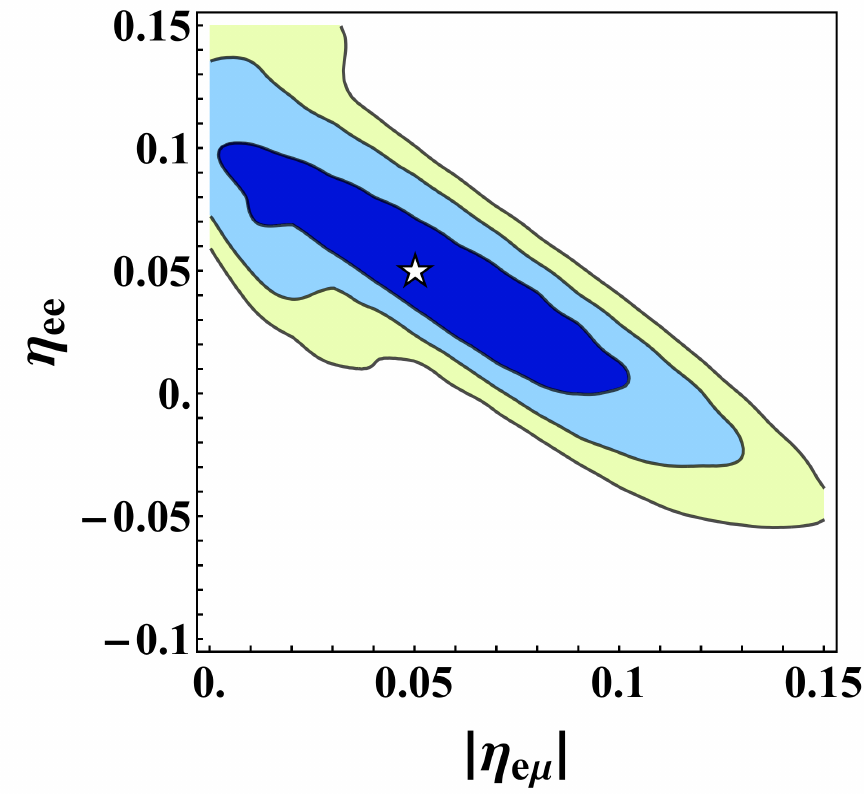}
    \includegraphics[width=0.328\linewidth, height=5cm]{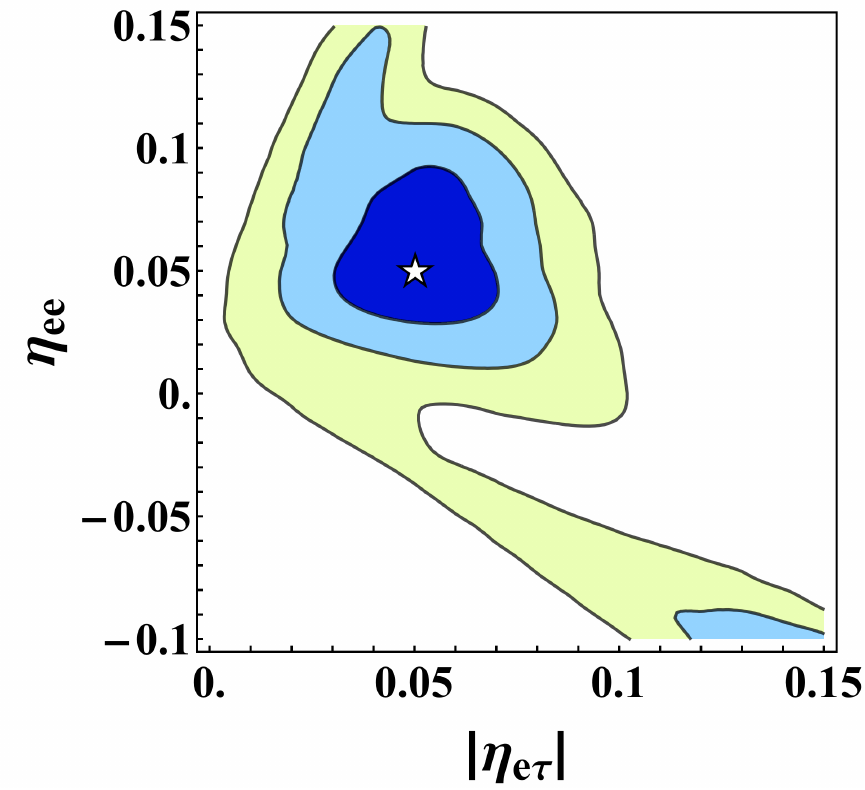}
    \includegraphics[width=0.328\linewidth, height=5cm]{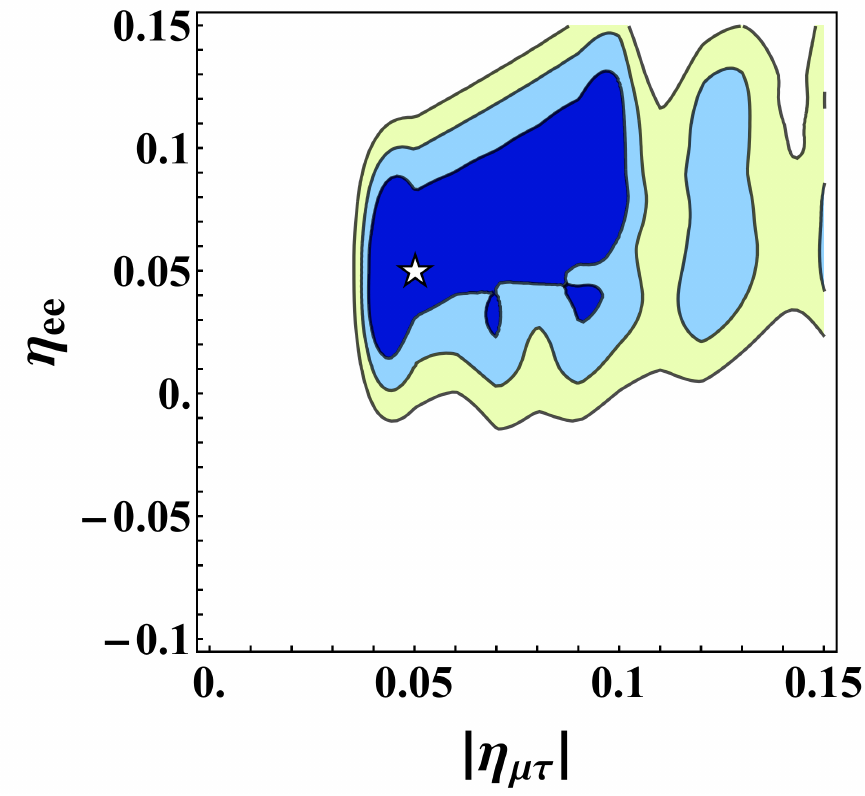}
    \includegraphics[width=0.328\linewidth, height=5cm]{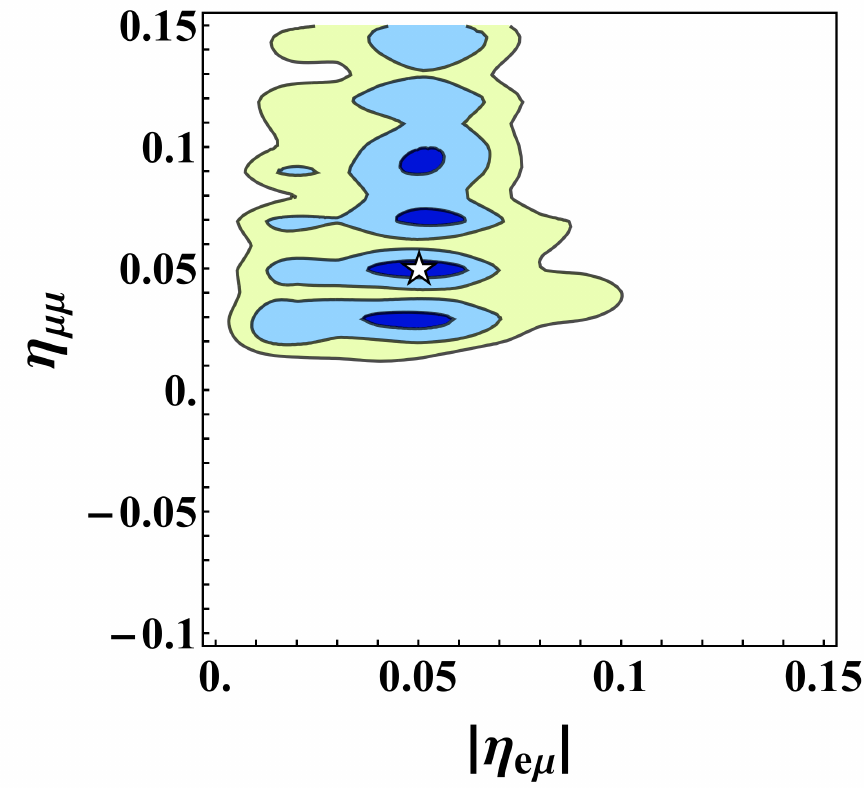}
    \includegraphics[width=0.328\linewidth, height=5cm]{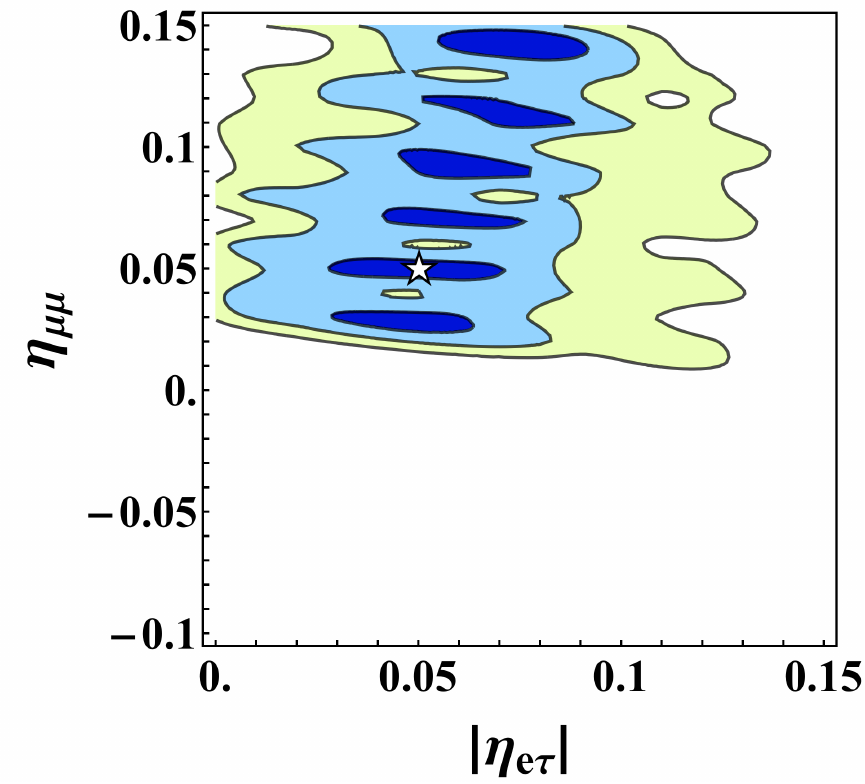}
    \includegraphics[width=0.328\linewidth, height=5cm]{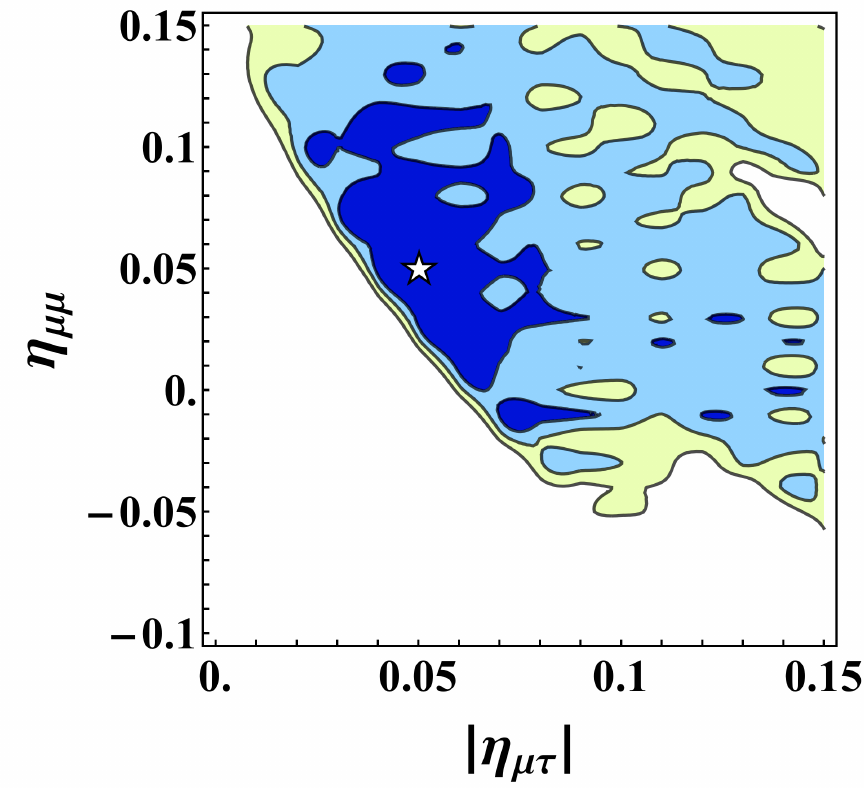}
    \includegraphics[width=0.328\linewidth, height=5cm]{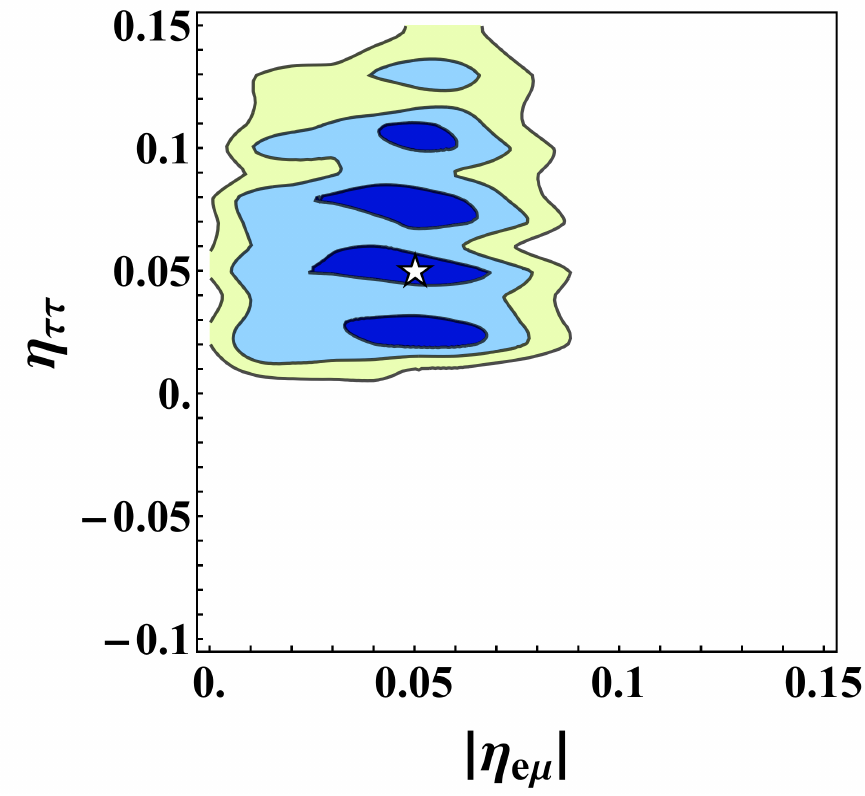}
    \includegraphics[width=0.328\linewidth, height=5cm]{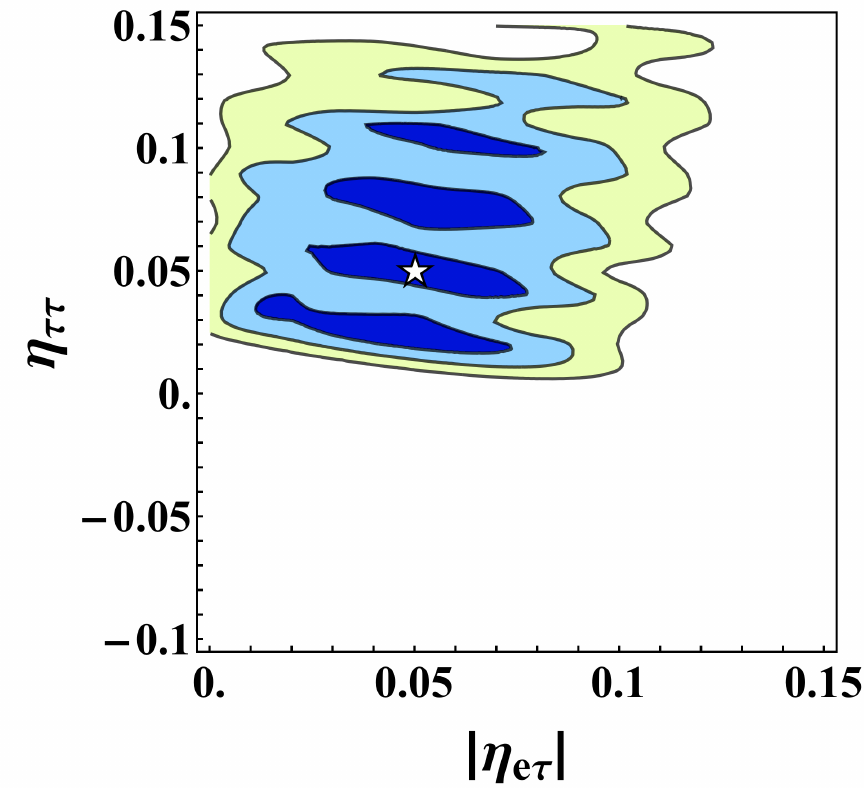}
    \includegraphics[width=0.328\linewidth, height=5cm]{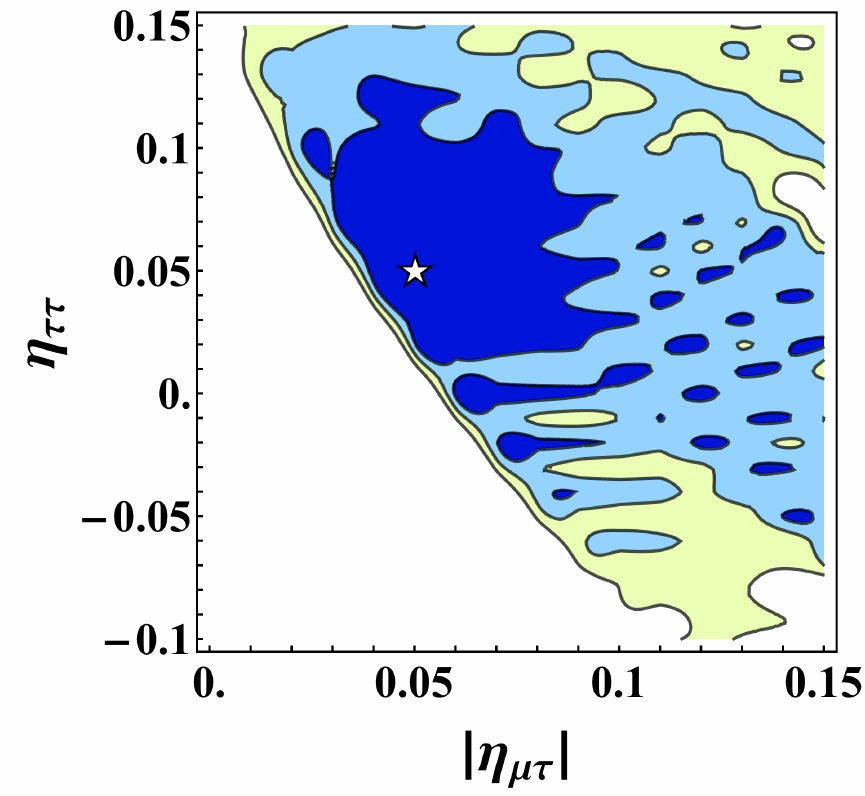}
    \caption{Correlation between diagonal ($\eta_{ee}$,$\eta_{\mu\mu}$,$\eta_{\tau\tau}$) and off-diagonal ($|\eta_{e\mu}|$,$|\eta_{e\tau}|$,$|\eta_{\mu\tau}|$) scalar NSI elements at DUNE. The blue, cyan and green colors represent the $1\sigma$, $2\sigma$ and $3\sigma$ allowed regions, respectively. The best-fit point is represented by the white solid star.}
    \label{fig:corr_diag_offdiag}
\end{figure}

In this section, we explore the correlation between different scalar NSI elements, taking the DUNE experiment as a case study. To study the correlation, we consider two non-zero scalar NSI parameters at a time out of the diagonal ($\eta_{ee}$,$\eta_{\mu\mu}$,$\eta_{\tau\tau}$), off-diagonal ($|\eta_{e\mu}|$,$|\eta_{e\tau}|$,$|\eta_{\mu\tau}|$) and associated phases ($\phi_{e\mu}$,$\phi_{e\tau}$,$\phi_{\mu\tau}$). We have explored four different scenarios: a) both elements are diagonal; b) both elements are off-diagonal; c) one element is diagonal and the other is off-diagonal; d) one element is off-diagonal $(\eta_{\alpha\beta})$ and the other is the associated phase $(\phi_{\alpha\beta})$. In all cases, we have plotted the allowed regions at $1\sigma$, $2\sigma$ and $3\sigma$ C.L. on the two-dimensional planes. The values of the oscillation parameters used for the simulation are as described in table \ref{tab:param}. We have chosen the normal mass ordering as the true mass ordering and higher $\theta_{23}$ octant as the true octant.

\begin{figure}[!h]
    \centering
    \includegraphics[width=0.7\linewidth]{new_figs/eta_corr_plots/legend.pdf}
    \includegraphics[width=0.328\linewidth, height=5cm]{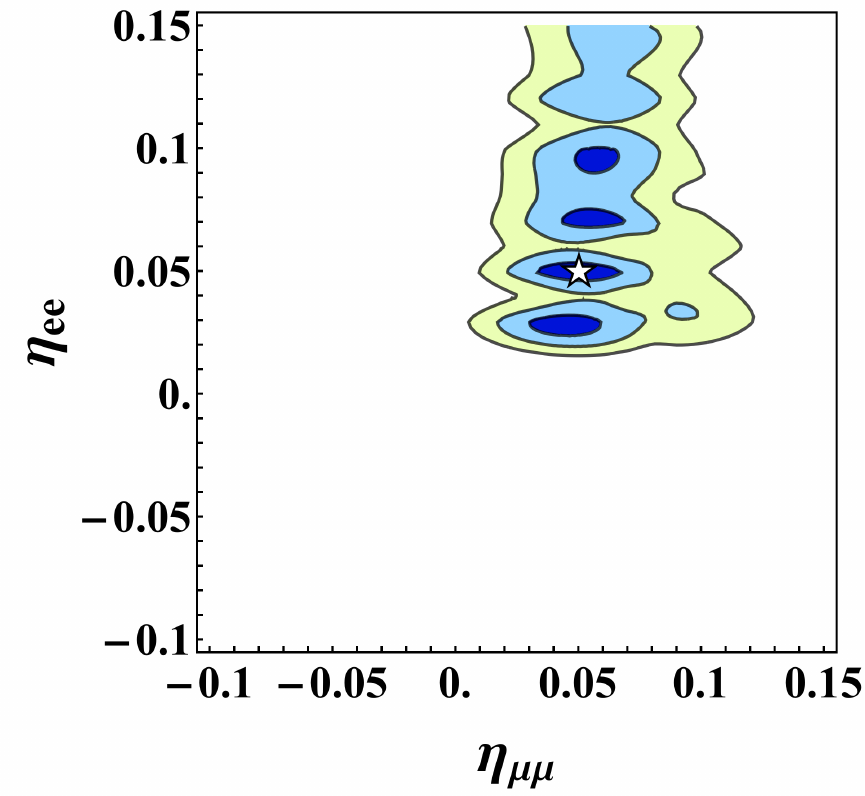}
    \includegraphics[width=0.328\linewidth, height=5cm]{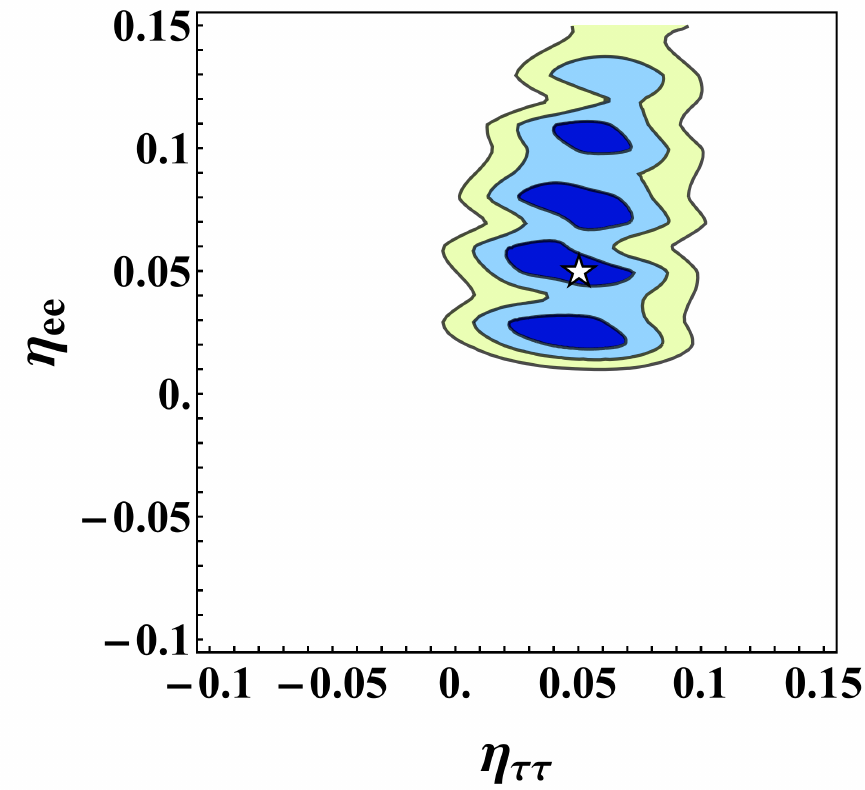}
    \includegraphics[width=0.328\linewidth, height=5cm]{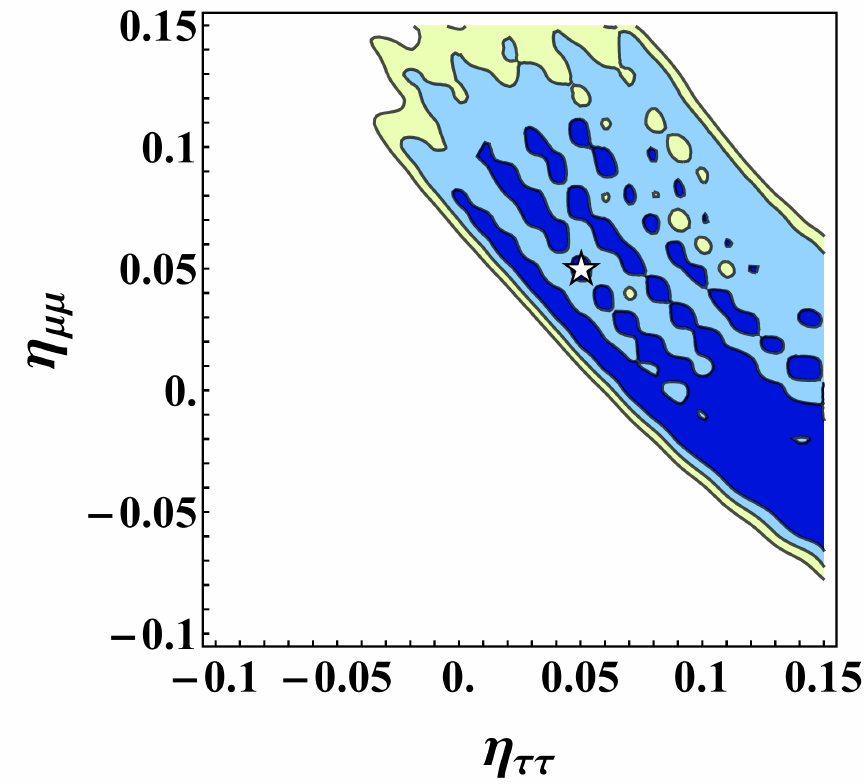}
    \includegraphics[width=0.328\linewidth, height=5cm]{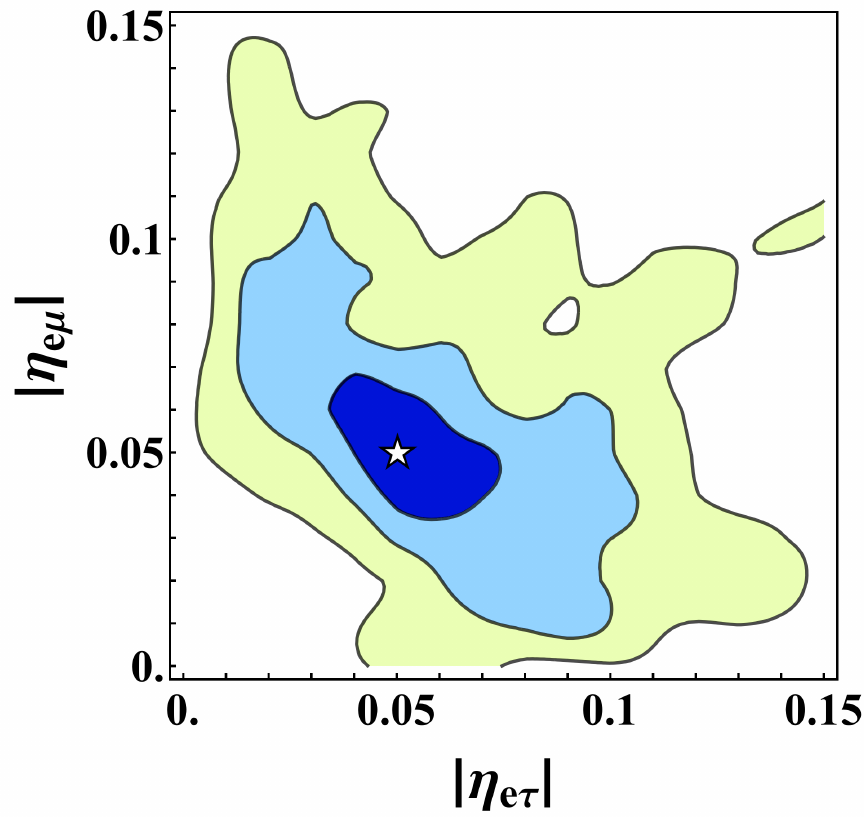}
    \includegraphics[width=0.328\linewidth, height=5cm]{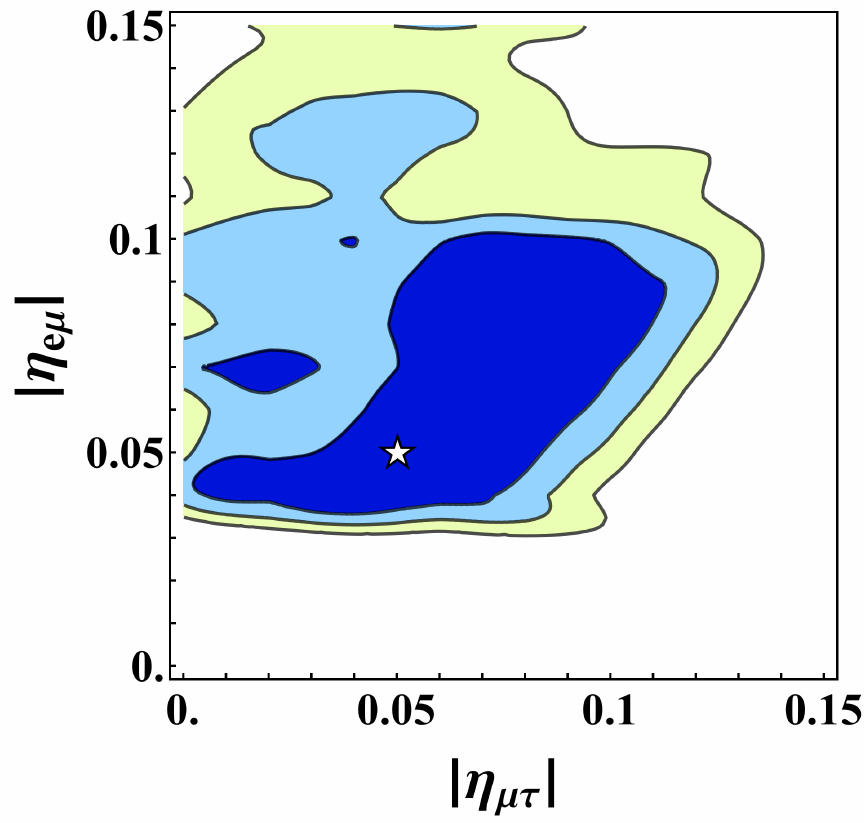}
    \includegraphics[width=0.328\linewidth , height=5cm]{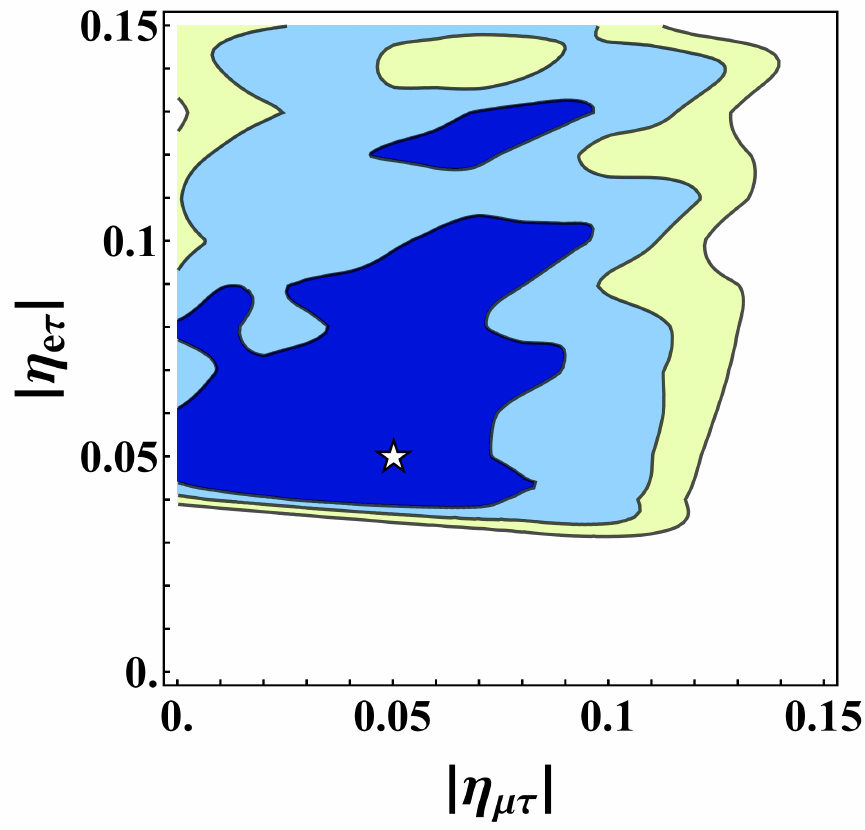}
    \includegraphics[width=0.328\linewidth, height=5cm]{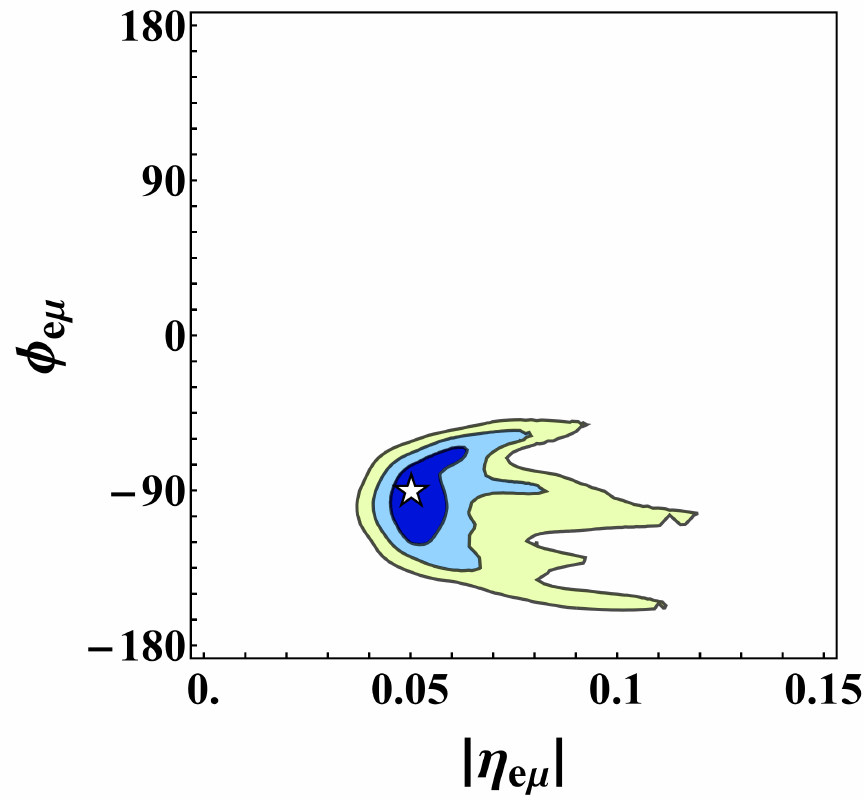}
    \includegraphics[width=0.328\linewidth, height=5cm]{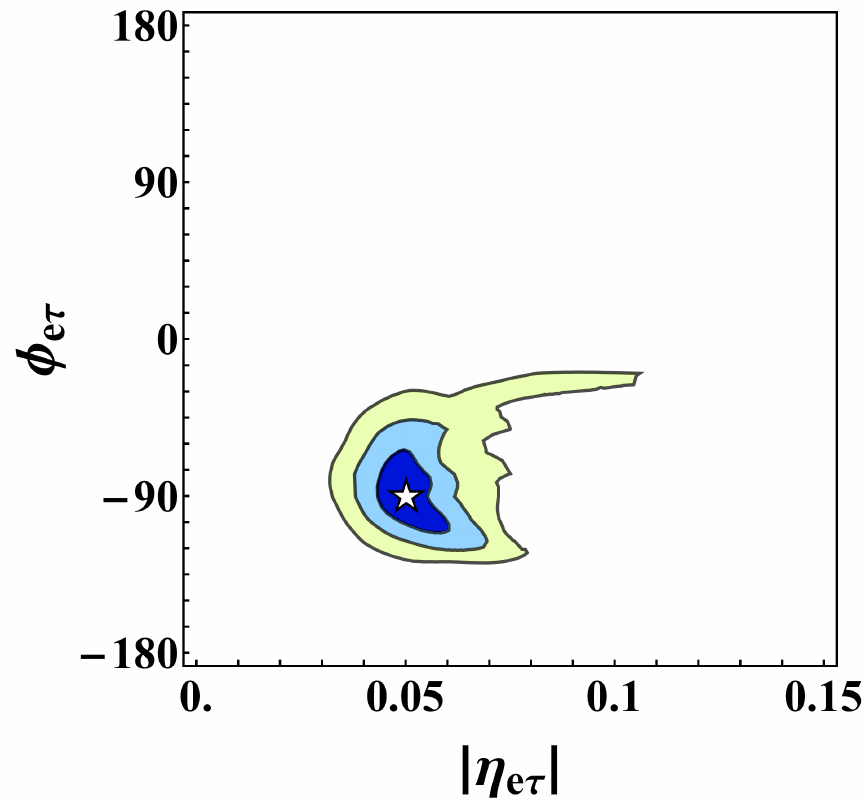}
    \includegraphics[width=0.328\linewidth, height=5cm]{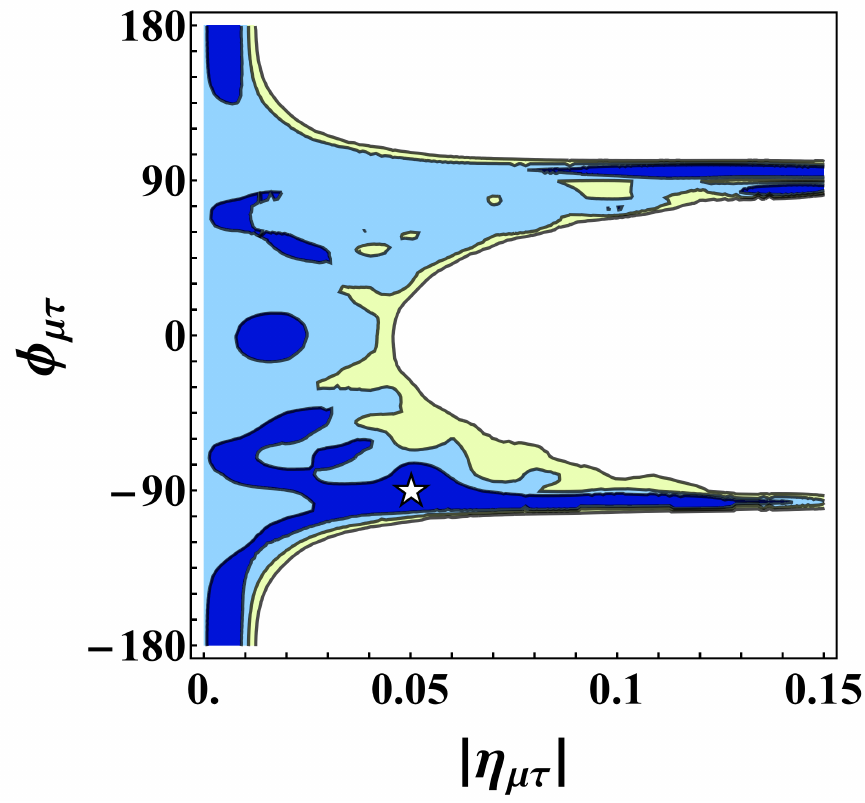}
    \caption{\textbf{Top panel}: Correlation among the diagonal elements ($\eta_{ee}$, $\eta_{\mu\mu}$, $\eta_{\tau\tau}$) only.\\ \textbf{Middle panel}: Correlation among the off-diagonal elements ($|\eta_{e\mu}|$, $|\eta_{e\tau}|$, $|\eta_{\mu\tau}|$) only.\\ \textbf{Bottom panel}: Correlation among the off-diagonal elements ($|\eta_{e\mu}|$, $|\eta_{e\tau}|$, $|\eta_{\mu\tau}|$) and phases ($\phi_{e\mu}$, $\phi_{e\tau}$, $\phi_{\mu\tau}$). The blue, cyan and green colours represent the $1\sigma$, $2\sigma$ and $3\sigma$ allowed regions, respectively. The white solid star represents the best-fit point.}
    \label{fig:corr_diag_offdiag_only}
\end{figure}

In figure \ref{fig:corr_diag_offdiag}, we have explored the correlation between the diagonal ($\eta_{ee}$,$\eta_{\mu\mu}$,$\eta_{\tau\tau}$) and off-diagonal ($|\eta_{e\mu}|$,$|\eta_{e\tau}|$,$|\eta_{\mu\tau}|$) elements. We have marginalized over the oscillation parameters $\Delta m_{31}^{2}$, $\theta_{23}$ and the associated scalar NSI phases $\phi_{\alpha\beta}$. We then obtain the minimum $\Delta \chi^{2}$ by varying the scalar NSI parameters in the fit data. We have plotted the diagonal elements in the vertical axes and varied the values of the parameters in the range [-0.1,0.15]. We varied the off-diagonal elements in the horizontal axes in the range [0,0.15]. For each case, the true value of the parameters is fixed at (0.05,0.05). The observations are listed below as
\begin{itemize}
    \item In the top-left panel, we observe a correlation between the elements $\eta_{ee}$ and $|\eta_{e\mu}|$. We see a similar correlation between $\eta_{ee}$ and $|\eta_{e\tau}|$. 
    \item We also observe a weaker but similar correlation of the element $|\eta_{\mu\tau}|$ with $\eta_{\mu\mu}$ and $\eta_{\tau\tau}$. For all the other parameters, we do not observe any notable correlation. The multiple degenerate regions seen at $1\sigma$ regions are due to the marginalization of the oscillation parameters.
\end{itemize}

In top panels of figure \ref{fig:corr_diag_offdiag_only}, we have explored the correlation among the diagonal elements ($\eta_{ee}$,$\eta_{\mu\mu}$,$\eta_{\tau\tau}$) only. For the diagonal element case, we marginalize only over the oscillation parameters $\Delta m_{31}^{2}$ and $\theta_{23}$. In middle panels of figure \ref{fig:corr_diag_offdiag_only}, we have examined the correlation among the off-diagonal elements ($|\eta_{e\mu}|$,$|\eta_{e\tau}|$,$|\eta_{\mu\tau}|$) only. For the off-diagonal case, we have additionally marginalized over the associated phases $\phi_{\alpha\beta}$. For each scenario, the true values of the scalar NSI parameters are fixed at (0.05,0.05). 
\begin{itemize}
    \item We observe a correlation between the diagonal elements $\eta_{\mu\mu}$ and $\eta_{\tau\tau}$ as shown in the top-right panel. We do not see any strong correlation among the off-diagonal elements, as shown in the middle panels of figure \ref{fig:corr_diag_offdiag_only}.
\end{itemize}

In bottom panels of figure \ref{fig:corr_diag_offdiag_only}, we have explored the correlation between off-diagonal ($|\eta_{e\mu}|$,$|\eta_{e\tau}|$,$|\eta_{\mu\tau}|$) elements and their corresponding phases ($\phi_{e\mu}$,$\phi_{e\tau}$,$\phi_{\mu\tau}$). The true values are fixed at $(\phi_{\alpha\beta},|\eta_{\alpha\beta}|)\sim(-90^{\circ},0.05)$.

\begin{itemize}
    \item In the case of $\phi_{\mu\tau}$ and $\eta_{\mu\tau}$, we observe a degeneracy for $\phi_{\mu\tau}\sim \pm90^{\circ}$ as shown in the bottom-right panel. No significant correlation is seen for the other parameters.
\end{itemize}

\section{Summary and Conclusion}\label{sec:summary}
In this work, we have considered the upcoming long-baseline experiment DUNE which has a baseline of 1300km with a fiducial volume of 40kton to be situated at SURF. It will have a peak energy of around 2.5 GeV. It offers very high sensitivity to the neutrino mass ordering, leptonic CP phase and the octant of $\theta_{23}$. We have discussed the capability of this upcoming long-baseline experiment to determine the leptonic CP phase in the presence of scalar non-standard interactions. We first discuss how the presence of off-diagonal scalar NSI elements and their associated phases can deviate the probabilities from the standard interaction scenario. We find that the impact of $|\eta_{e\mu}|$ and $|\eta_{\mu\tau}|$ on $P_{\mu e}$ channel is significant. However, for a given value of $|\eta_{\alpha\beta}|$, the phases $\phi_{e\mu}$ and $\phi_{e\tau}$ have a greater impact on the probabilities in comparison to $\phi_{e\mu}$. We illustrate the differences in the probabilities by defining a quantity $\rm \Delta P_{\mu e}$ in the two-dimensional $(\eta_{\alpha\beta}$-$\delta_{CP})$ plane at the first oscillation maximum. In presence of $\eta_{e\mu}$ and $\eta_{e\tau}$, we find interesting degenerate regions at the probability level. The impact of $\eta_{\mu\tau}$ on $\Delta P_{\mu e}$ is relatively low. 

The presence of additional phases $\phi_{\alpha\beta}$ can further complicate the measurement capability of $\delta_{CP}$ at DUNE. We have explored the $\delta_{CP}$-constraining capability in the presence of off-diagonal scalar NSI elements $|\eta_{\alpha\beta}|$ and corresponding phases $\phi_{\alpha\beta}$. We observe that the constraining is much better in the presence of $|\eta_{\mu\tau}|$ compared to $|\eta_{e\mu}|$ and $|\eta_{e\tau}|$. The presence of phases $\phi_{e\mu}$ and $\phi_{e\tau}$ shows similar constraining of the leptonic CP-phase. However, we observe a degeneracy in the presence of scalar NSI phase $\phi_{\mu\tau}$ around $\pm90^{\circ}$. We have also explored the impact of scalar NSI on the CPV sensitivities at DUNE. We find that the phases $\phi_{e\mu}$ and $\phi_{e\tau}$ can affect the sensitivities strongly in comparison to $\phi_{\mu\tau}$. We have also explored how the absolute neutrino mass scale can affect the CPV sensitivities. We observe an overall enhancement with the increase in the lightest neutrino mass value. We also constrain the off-diagonal scalar NSI elements at $5\sigma$ CL. We observe that the bounds on the elements $|\eta_{e\mu}|$ and $|\eta_{e\tau}|$ improves significantly for higher neutrino mass scale. We have also examined the correlation of the scalar NSI parameters across different elements and phases, revealing correlations among various parameters. In particular, we find that the correlation between $|\eta_{\mu\tau}|$ and $\phi_{\mu\tau}$ leads to the degenerate regions observed in the $\delta_{CP}$ measurement. This study has shed light on the impact of off-diagonal elements and additional phases on the CP-measurement sensitivities at DUNE.

\section*{Acknowledgments}
The authors acknowledge the DST SERB grant CRG/2021/002961. AS acknowledges the CSIR SRF fellowship (09/0796(12409)/2021-EMR-I) received from CSIR-HRDG. DB would thank DST
INSPIRE Fellowship (DST/INSPIRE Fellowship/2022/IF220161) for providing financial
support. AM acknowledge the support of DST SERB grant  xPHYSPNx
DST00627xxBB006-ZBSA-3237 titled ``Indian Institutions-Fermi lab Collaboration in Neutrino Physics" for the financial support.

\bibliographystyle{JHEP}
\bibliography{main}

\end{document}